\documentclass{agujournal2019_ArXiv}
\usepackage{url} 
\usepackage[inline]{trackchanges} 
\usepackage{soul}

\usepackage{amsmath}
\usepackage{amssymb}

\definecolor{vertdc1}{RGB}{20,89,33}
\definecolor{blood}{RGB}{193,41,41}
\definecolor{viol}{RGB}{109,10,186}
\definecolor{dgreen}{RGB}{9,95,29}
\definecolor{dorange}{RGB}{197,69,6}

\definecolor{CNRSBlue}{RGB}{26,48,81}
\definecolor{CNRSLightBlue}{RGB}{10,141,167}
\definecolor{DarkBlue1}{RGB}{13,7,136}
\definecolor{DarkRed}{RGB}{212,0,0}
\definecolor{DarkRed2}{RGB}{150,0,0}
\definecolor{Violet}{RGB}{122,47,214}
\definecolor{DarkGreen}{RGB}{21,91,16}

\newcommand{\rev}[1]{#1}

\draftfalse

\journalname{JGR: Planets}

\begin{document}
\nolinenumbers

%
%


\title{Capillary processes in extraterrestrial contexts}

%
%




\authors{Daniel Cordier\affil{1}, G\'{e}rard~Liger-Belair\affil{1}, 
        David~A.~Bonhommeau\affil{2},
        Thomas~S\'{e}on\affil{3},
        Thomas~App\'{e}r\'{e}\affil{4},
        and
        Nathalie~Carrasco\affil{5} }

\affiliation{1}{Universit\'{e} de Reims Champagne Ardenne, CNRS, GSMA UMR 7331, 51100 Reims, France}
\affiliation{2}{Universit\'{e} Paris-Saclay, Univ Evry, CY Cergy Paris Universit\'{e}, CNRS, LAMBE, 91025, Evry-Courcouronnes, France}
\affiliation{3}{Sorbonne Universit\'{e}, CNRS, UMR 7190, Institut Jean Le Rond d'Alembert, F-75005 Paris, France}
\affiliation{4}{Lyc\'{e}e Saint-Paul, 12, all\'{e}e Gabriel Deshayes, 56000 Vannes, France}
\affiliation{5}{LATMOS, UMR CNRS 8190, Universit\'{e} Versailles St Quentin,
                 UPMC Univ. Paris 06, 11 blvd d'Alembert,
                 78280 Guyancourt, France}
                 




\correspondingauthor{Daniel Cordier}{daniel.cordier@cnrs.fr}
%


\begin{keypoints}
\item Capillarity is involved in numberless natural processes.
\item During past decades, the presence of liquid phase of different nature has been detected on 
      several extraterrestrial bodies of the solar system.
\item We review and discuss physical processes that may have important consequences in planetary science like
      the transport of organic material between \rev{planetary interior and the atmosphere}.
\end{keypoints}

%
%

%



\begin{abstract}
   The Earth is no longer the only known celestial body containing one or more
   liquid phases. The Cassini spacecraft has discovered seas of hydrocarbons at the surface of Titan, while 
   a series of corroborating evidences argue in favour of the existence of an aqueous ocean beneath the icy crust of 
   several moons.
   Capillarity  embraces a family of physical processes occurring at the free surface of
   a liquid. These phenomena depend on the liquid properties and on the local planetary conditions.
   Capillarity may have important direct or indirect implications on the geoscientific and astrobiological points of view.
   In this paper, we discuss capillarity physics among solar system objects and expected consequences for planetary
   science.
\end{abstract}

\section*{Plain Language Summary}

  The formation of raindrops, the production of tiny liquid droplets by bubble bursting at the 
surface of an ocean, or even the floatation of small solid particles at this surface, are examples of
\rev{capillarity effects, physical processeses which can be} observed in everyday life. In this work, we discuss such phenomena 
in \rev{extraterrestrial} contexts where large amount of liquids have be detected during past decades by space exploration.

%
%

%


%
%
%
%

\section{\label{intro}Introduction}

\begin{figure}[]
\nolinenumbers
\begin{center}
\includegraphics[angle=0, width=7 cm]{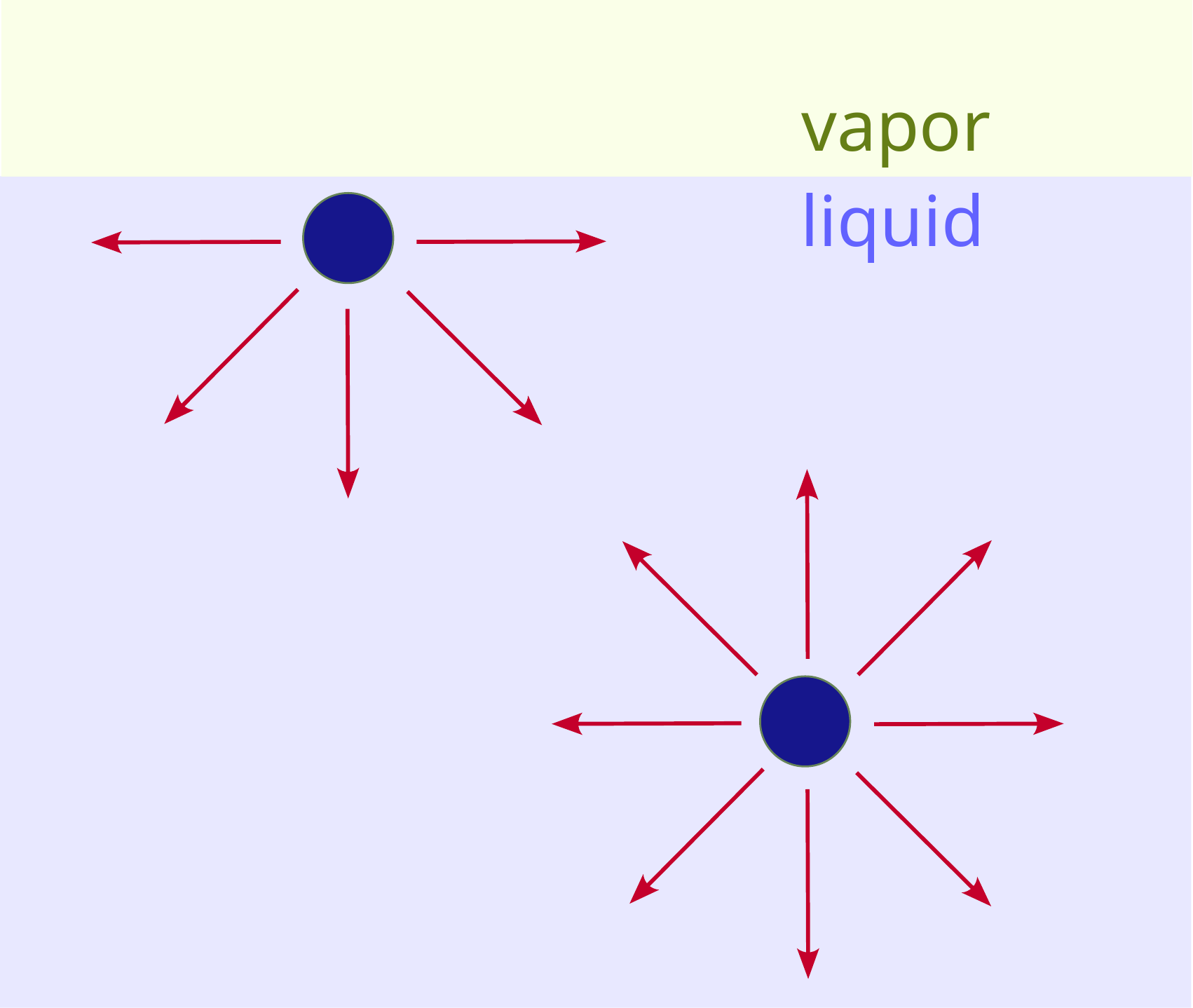}
\caption[]{\label{MoleInteract} A schematic view of a liquid-vapor interface, two ``molecules'' are represented by
           blue circles, while the intermolecular interactions are symbolized by red arrows.}
\end{center}
\end{figure}

\begin{figure}[]
\begin{center}
\includegraphics[angle=0, width=8 cm]{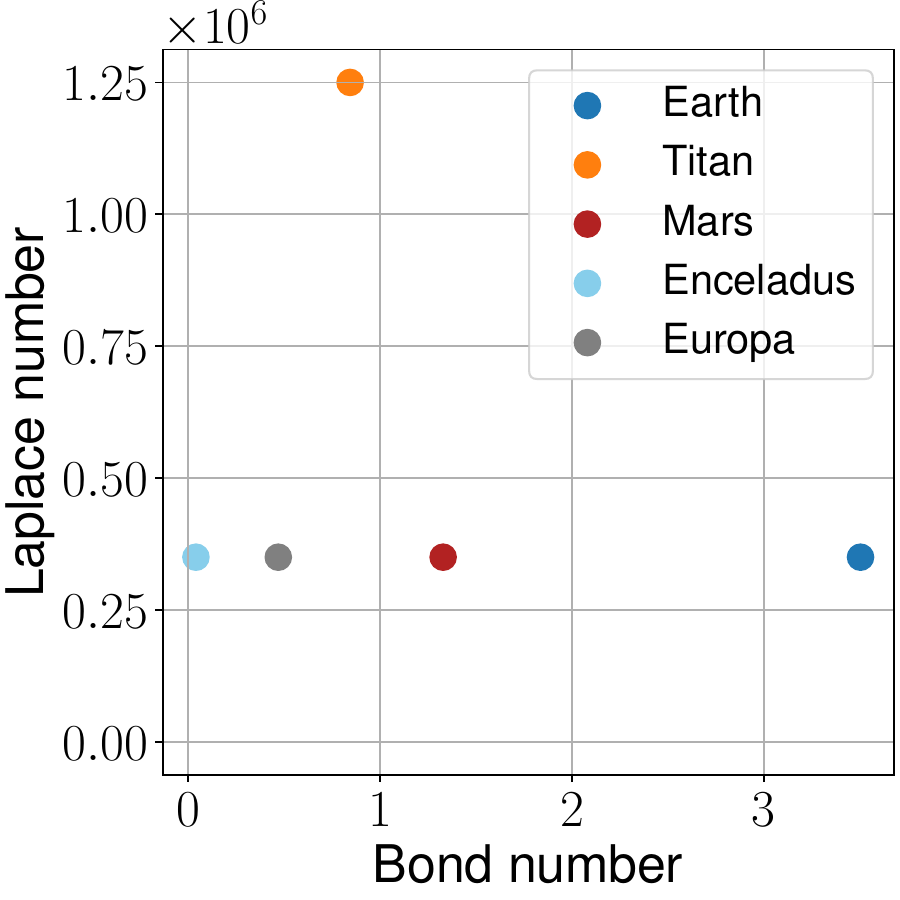}
\caption[]{\label{BoLa}The Laplace number as a function of the Bond number for the planetary bodies considered
           in this article. The considered characteristic length $L$, required in both $La$ and $Bo$ computations, is the same for 
           all bodies, and arbitrary fixed to 5~mm.}
\end{center}
\end{figure}

  In everyday life, \rev{capillary action manifests itself in many different ways}, going from liquid ascension in fibers to the 
dynamics of bubbles and droplets. On a scientific point of view, hydrodynamics of capillarity is a rather old field, but it is still 
the subject of very active researches \cite{degennes_etal_2004,drelich_etal_2020} with a wide variety of applications like 
the manufacture of micro-lenses or the understanding of blood circulation.
\rev{In all these physical processes, we always find at least one interface between two immiscible
liquids, or a liquid and a vapor. We also have situations involving three phases: two liquids and a solid; this is the case in problems 
related to surface wetting. These interfaces have particular hydrodynamic behaviors, this specificity is fundamentally governed by
different states of molecules in the vicinity
of the interface (see Fig.~\ref{MoleInteract}) compared to molecules in the midst of the 
liquid \cite{degennes_etal_2004}. As shown in Fig.~\ref{MoleInteract}, this dichotomy is explained by the loss of half of cohesive 
interactions by molecules located at the surface.}\\
%
%
   In this work, we investigate the role of capillarity in some phenomena relevant to planetary contexts.
The two last decades of space exploration have revealed the existence of several extraterrestrial liquid phases: 
liquid methane-ethane-nitrogen mixtures in Titan's polar regions \cite{stofan_etal_2007}, liquid water for Enceladus \cite{porco_etal_2006} 
and probably Europa \cite{roth_etal_2014,sparks_etal_2017,roth_2021}, while many evidence speak in favor of the massive presence of 
liquid water at the surface of Mars, 3 to 4 billions years ago \cite{nazari-sharabian_etal_2020}.\\
   In order to compare the relative importance of capillarity and gravity, physicists have introduced the Bond number $Bo$ 
(also called E\"{o}tv\"{o}s number) \cite{seon_2018}
\begin{equation}\label{BondNumb}
 Bo = \frac{\rho_{\rm liq} g L^2}{\sigma}
\end{equation}
with $\rho_{\rm liq}$ the density of the liquid (kg~m$^{-3}$), $g$ the gravity (m~s$^{-2}$), $L$ the typical size (m) of a considered
object, and $\sigma$ the surface tension (N~m$^{-1}$). A large Bond number, {\it i.e.} $Bo \gg 1$, is the signature of a physical 
process where gravity has a dominant role; in the opposite situation capillarity is prominent. Clearly, for the same liquid at the surface 
of celestial bodies where gravity is smaller than the terrestrial one, the effects of capillarity may be significantly reinforced. 
If we consider the capillary length $l_{c} = \sqrt{\sigma/\rho g}$ ({\it cf}. p.~33 \citeA{degennes_etal_2004}), we see that effects of capillarity should be
important for larger objects under lower gravity. For instance, the small kronian satellite, Enceladus, has a crust dotted by cracks 
\cite{postberg_etal_2018}, where the presence of liquid water is suspected. The influence of capillarity should be strong there, compared with
what we have at the surface of the Earth, since the gravity of Enceladus is around $0.11$~m s$^{-2}$, almost two orders of magnitude lower than 
the terrestrial value (9.81~m~s$^{-2}$). 
\rev{By analogy with Reynolds number, we can also define the Laplace number $La= \rho\sigma L /\eta^2$ (where $\eta$ is the dynamic viscosity)
which quantifies the ratio between the inertio-capillary effect and and viscous effects \cite{deike_etal_2018}. In Fig.~\ref{BoLa} we have reported the behavior
of $Bo$ {\it versus} $La$ for different planetary objects. Bodies where liquid water exists, are aligned at $La \simeq 0.3 \times 10^6$ and Titan, harboring 
liquid methane, stands clearly apart. This figure shows that Enceladus and Titan should exhibit the most striking differences with the Earth.}\\
Our perspective
is different from the one adopted in studies based on experiments performed under microgravity conditions
({\it i.e.} aboard the International Space Station, \citeA{weilogel_etal_2009}).
In the latter case, the purpose is focused on the fundamental properties of capillary hydrodynamics, while we are discussing applications 
in the particular field of planetology.\\
%
%
In the following, we first address the problem of small solid particles floatation at the surface
of a liquid, and we discuss the aerosol formation via bubble bursting through these surfaces. 
After these aspects focused on liquid/gas interface dynamics, we turn our attention to raindrop size distribution and 
expected consequences in extraterrestrial contexts. A particular focus is on possible observations by {\it Dragonfly}, 
the future space mission to Titan, that is planned to explore the surface of this satellite of Saturn by the mid-2030s.

\section{Floatability by Capillarity}
\label{floata_capill}
\begin{figure}[]
\begin{center}
\includegraphics[angle=0, width=6 cm]{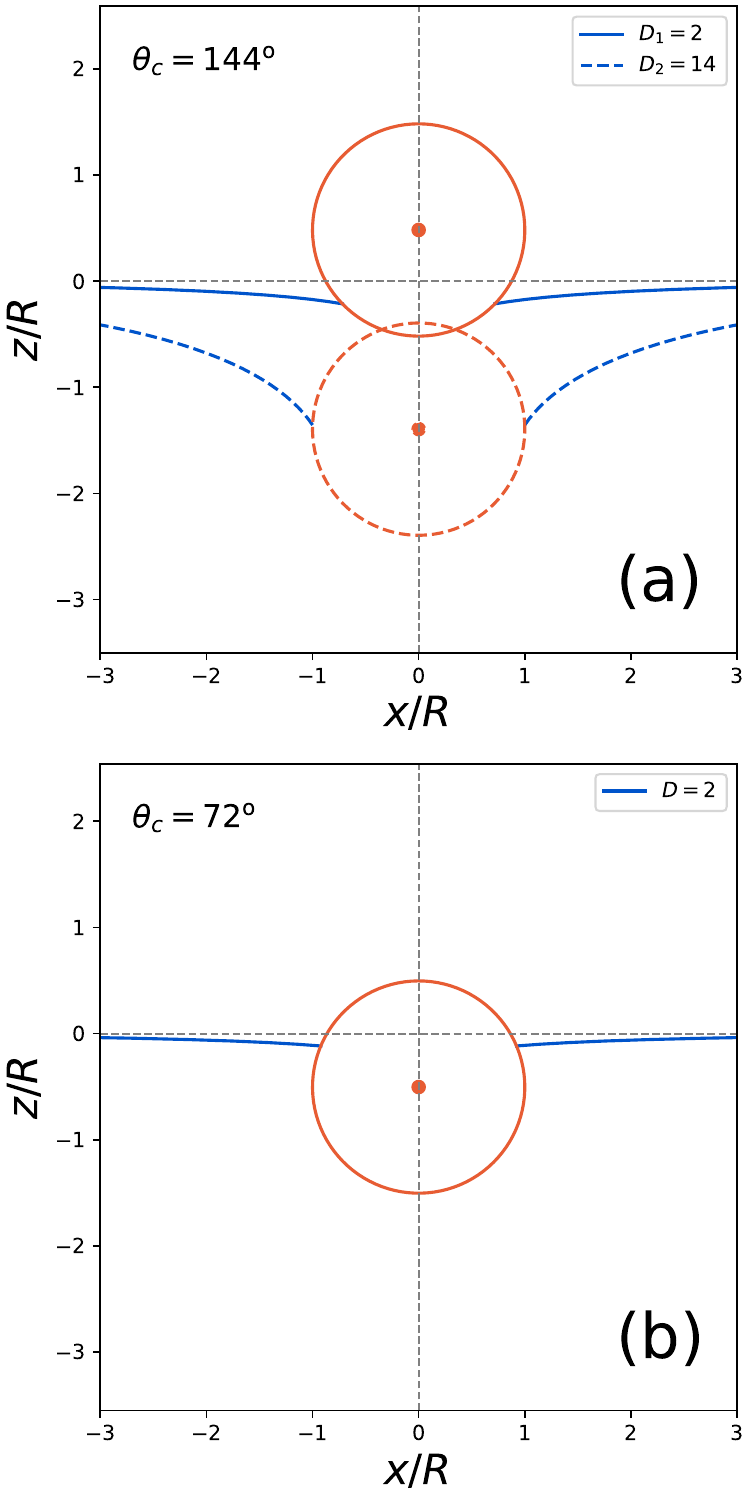}
\caption[]{\label{influParam}Meniscus profiles (in blue) due to a solid spherical particle (in orange)
           obtained numerically using the method described by \citeA{lee_2018}. 
           The coordinates $z$ and $x$ have been normalized using the particle radius $R$.
           All solutions have the same Bond number $\mathcal{B}o= 0.1$ (defined by Eq.~\ref{BondNumb}). (a) The influence of the 
           density ratio 
           $D = (\rho_{\rm sol} - \rho_{\rm air}) / (\rho_{\rm liq} - \rho_{\rm air})$, for a contact angle $\theta_c$ fixed to 
           $144^{\rm o}$.
           (b) An example of flotation of a liquidophobic sphere with $\theta_c= 72^{\rm o}$, this result corresponds to $D= 2$.}
\end{center}
\end{figure}
%
       The presence of some floating material, over an interface between a liquid phase and a gaseous phase, can lead to
major consequences by affecting exchanges of energy, momentum and matter between the phases. The form of this floating material may 
vary from a tiny monomolecular microlayer to a thick layer similar to ice pack. At the surface of terrestrial oceans, the presence of 
a thin floating film, produced by biological activity \cite{lin_etal_2003}, has a damping effect on wave activity; 
marine films is an important topic, and constitutes a full-fledged academic discipline \cite{marine_surface_films_2006}.
Broadly speaking, a solid material is able to float at a liquid surface with the help of two physical processes: (1) the well known 
Archimede's buoyancy force, for materials less dense than the liquid, (2) capillary forces resulting from interaction at molecular scale. 
This is the second effect we are studying in this 
paragraph. 

The floatability generated by capillarity has been studied for a long time since it is essential for industrial applications
based on floatation \cite{chipfunhu_etal_2011,mousumi_venugopal_2016,kyzas_etal_2018}. For the sake of simplicity, classical studies of 
solid particles floatability consider idealized spherical particles \cite{scheludko_etal_1976,crawford_ralston_1988} and 
we maintain this approach here (see Fig.~\ref{Fig1Lee18} in \ref{AppCondi}). The differential equation provided by 
first principles can be solved numerically (see~\ref{AppCondi}), the main input parameters being the particle radius $R$, the 
planetary gravity $g$, the liquid surface tension $\sigma$, the liquid-solid contact angle $\theta_c$, and the respective densities 
of the liquid ($\rho_{\rm liq}$) and solid ($\rho_{\rm sol}$). Examples of results as a function of the meta-parameter 
$D=(\rho_{\rm sol}-\rho_{\rm air})/(\rho_{\rm liq}-\rho_{\rm air})$ can be seen in Fig.~\ref{influParam}. As expected, denser solids 
lead to the most incurved menisci. The liquid surface tension has the opposite effect, with almost flat liquid-air interface for high 
surface tension values. In Fig.~\ref{influParam}-b the effect of a relatively small contact angle $\theta_c$ is represented. This case shows 
clearly the possibility for floatation, even in the case of liquidophobic ({\it i.e.} $\theta_c < 90^{\rm o}$) solid material. 
However, in such a situation, the filling  angle
$\theta_f$ (see Fig.~\ref{Fig1Lee18} in~\ref{AppCondi}) tends to be small. It can be easily shown that (see~\ref{AppCondi})
that the radius of a floating spherical particle has a maximum radius $R_{\rm max}$
\begin{equation}\label{Rmax}
  R_{\rm max} \sim \sqrt{\frac{3\sigma}{2 (\rho_{\rm sol} - \rho_{\rm liq}) g}} \, \sin \frac{\theta_c}{2}
\end{equation}
%
\begin{figure}[]
\begin{center}
\includegraphics[angle=0, width=8 cm]{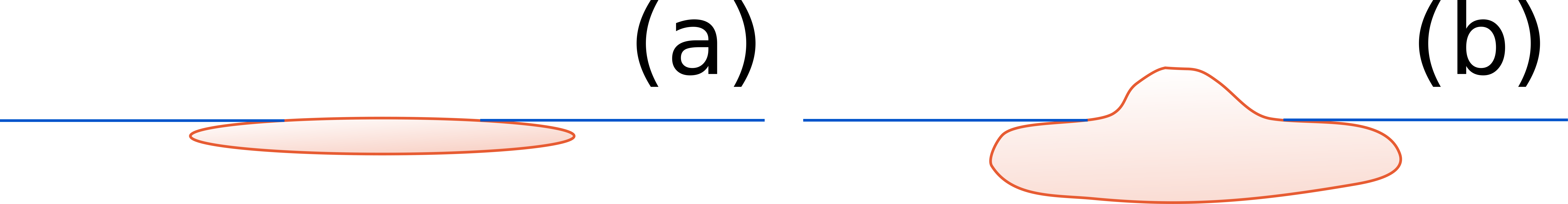}
\caption[]{\label{examples_oblate}Panel (a) and (b): examples of solid particle shapes (in orange) that could 
           float even with a very small contact angle.}
\end{center}
\end{figure}
%
This equation can be found in the literature in a simplified form 
(see Eq.~4.5 in \cite{scheludko_etal_1976}).
Not surprisingly, when $\rho_{\rm liq}$ tends to $\rho_{\rm sol}$ the radius $R_{\rm max}$ becomes arbitrarily large.
Thus, in principle, any small enough spherical particle can float for a non-zero contact angle with the constraint $\theta_f \lesssim \theta_c$.
The latter condition puts restrictions on the size of actual particles that are able to float. Up to this point, we have only considered 
spherical particles but the reality may be more complex. In Fig.~\ref{examples_oblate} we have represented examples of shapes that 
could favor floatability for materials with a small contact angle. 

Another aspect, ignored in the above lines, concerns the 
stability of the equilibrium involved in flotation \cite{huh_mason_1974}: is this equilibrium stable against translational or
rotational perturbations? Such a question remains very complicated to answer and is clearly beyond the scope of this paper.
Nonetheless, we emphasize that stability becomes an important issue when perturbations length scale, or ``wavelength'' if appropriate,
is comparable to, or smaller than, the size of the particle to efficiently affect the stability of the system. If the 
perturbation is a wave, the frequency $\nu$ depends on the  wavelength $\lambda$ according to the dispersion relation for 
capillary waves \cite{melville_2001}
\begin{equation}\label{capwavefreq}
\nu = \sqrt{\frac{2\pi \, \sigma}{\rho \lambda^3}}
\end{equation}
In conclusion, this discussion shows us that a large category of particles can float over a liquid-gas interface, even liquidophobic 
materials under some conditions concerning their shape, size and contact angle.

\subsection{The case of Titan}
%
\begin{figure}[]
\begin{center}
\includegraphics[angle=0, width=8 cm]{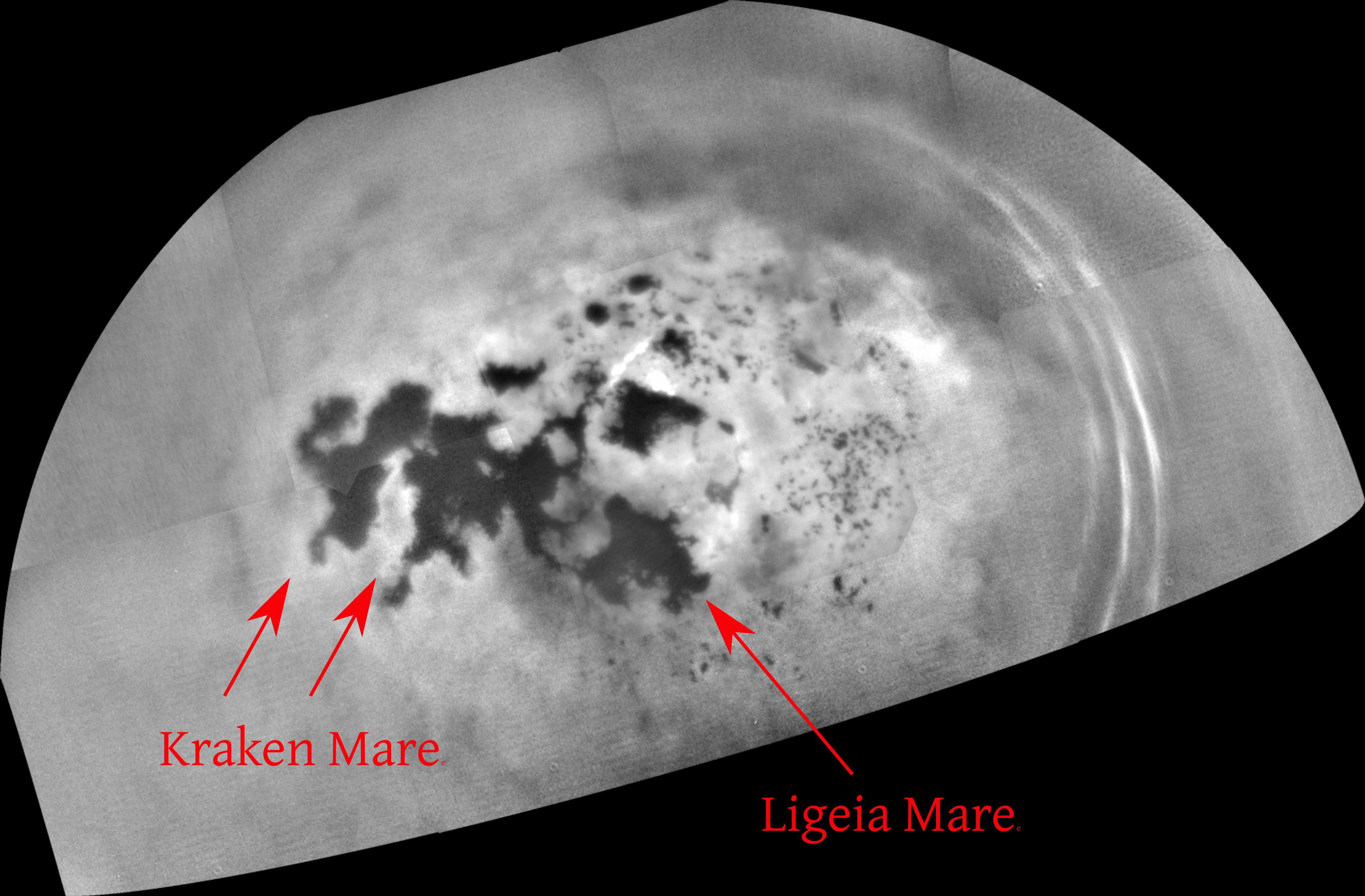}
\caption[]{\label{TitanLakes}A mosaic of infrared images of Titan north polar regions taken by the 
           Narrow Angle Camera of the ISS instrument aboard the Cassini spacecraft. The main hydrocarbon seas are clearly visible, their
           sizes are similar to those of the Great American lakes (Image Credit: NASA/JPL-Caltech/Space Science Institute).}
\end{center}
\end{figure}
%
The Cassini orbiter has revealed a collection of seas and lakes in the polar regions of Titan (see Fig.~\ref{TitanLakes}), 
the largest satellite of Saturn \cite{stofan_etal_2007}. Besides, Titan is also the only moon of the solar system possessing 
a dense atmosphere, which harbors a thick layer of organic haze. This unique feature has been the source of inspiration 
for many works regarding photochemistry and aerosol properties
\cite{mullerwodarg_etal_2014}. The potential floatability of these aerosols, at the surface of Titan's seas, has been
already discussed in an idealized, if not oversimplified, way \cite{cordier_carrasco_2019}, and may have an enhanced 
wave damping effect, that could explain the \rev{radar and near-infrared, smooth aspect of these liquid bodies 
\cite{stephan_etal_2010,barnes_etal_2011b,wye_etal_2009,zebker_etal_2014,grima_etal_2017,cordier_carrasco_2019}}.

Although the chemical composition of Titan's seas is only approximately known, reasonable estimates can be made, leading to 
a density around $500$~kg~m$^{-3}$ and a surface tension of about $\sim 20$~mN~m$^{-1}$ \cite{cordier_etal_2017a}. Laboratory 
determinations on Titan's aerosol analogs, called tholins \cite{sagan_etal_1992}, lead to densities around $\sim 0.8$~g~cm$^{-3}$ 
for high-pressure experiments \cite{horst_tolbert_2013}, and densities in the range $1.3$--$1.4$~g~cm$^{-3}$ for low-pressure simulations
\cite{imanaka_etal_2012,brouet_etal_2016}. However, for low-pressure measurements, densities can be found as small as 
$0.4$~g~cm$^{-3}$ \cite{horst_tolbert_2013}, 
a value for which materials could simply float due to buoyancy forces. 
The wettability of aerosols remains essentially poorly constrained.
Only a few derivations from measurements on laboratory analogs, are available. These determinations draw the picture of very 
wettable material with contact angle around $\theta_{\rm c}~=~5^{\rm o}$ \cite{rannou_etal_2019}. 
A contact angle of strictly 
$\theta_{\rm c}= 0^{\rm o}$ \cite{yu_etal_2020} is probably an idealization since it is not compatible with the conservation of 
mass principle. 
Titan's aerosols appear to be fractal aggregates of spherical 
monomers \cite{seignovert_etal_2017}, for such a single sphere the use of Eq.~\ref{Rmax} leads to $R_{\rm max} \sim  2 \times 10^{-4}$~m if
we adopt a particle density of $10^{3}$~kg~m$^{-3}$, together with Titan's gravity $g_{\rm Tit} = 1.352$~m~s$^{-2}$. 
This estimate is well above the monomer radius estimations of $\sim 50$~nm 
\cite{seignovert_etal_2017} enabling Titan's aerosol particles to float.
We stress that these particles 
may have a variety of properties (chemical surface properties, size polydispersity, diversity of shapes, etc), leading to various 
floatabilities. In addition, when the contact angle is very small ({\it e.g.} a few degrees) a spherical particle would be almost fully
immersed in the liquid, but according to Eq.~\ref{capwavefreq}, only perturbations at frequencies above $\sim 500$~MHz, which are unlikely, 
could destabilize the particle and make it sink.
Then, the sea surface could be regarded as a 
``filter'', trapping  only a fraction of atmospheric sediments.\\
   Finally, we recall that solid particles are not, by far, the only candidates for ``floating microlayer builders'' for Titan's 
seas. Molecules produced by chemical reactions in the Titan's atmosphere, like polyacetylenes
\cite{chien_1984,handbook_of_conducting_polymers_1998,elachi_etal_2005}, are favoured due to their low densities, while 
surfactant molecules like acrylonitrile \cite{stevenson_etal_2015a} may also \rev{form thin monomolecular layers over}
the surface of the seas.\\
\rev{In the titanian case, other aspects have been also explored. For instance, the hypotheses of floating ``pumices'', made of very 
porous organic material to ensure buoyancy, has been proposed \cite{yu_etal_2024}. Laboratory experiments have shown \cite{farnsworth_etal_2023} that 
non-coalescing liquid droplets may exist at the surface of methane-rich seas after hydrocarbon rainfalls. Interestingly, 
this phenomenon may appear to be independent of the liquid density. The floatation of ethane-rich drops is ensured in a wide range of sea bulk
liquid compositions, while floating methane enriched droplets only exist in narrow ranges of composition and impact velocity.}

%
\begin{figure}[]
\begin{center}
\includegraphics[angle=0, width=8 cm]{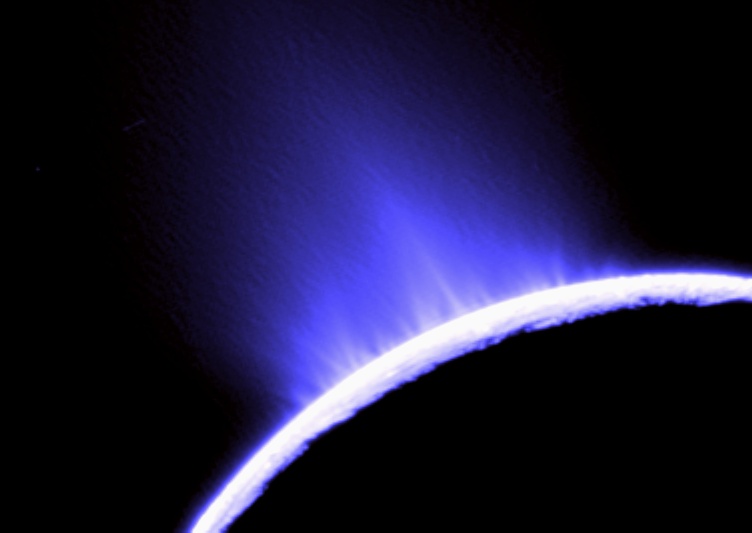}
\caption[]{\label{EncGeysers}The plumes of ejected gases and icy dust from the south pole of the tiny moon Enceladus,
           whose diameter is approximately $500$~km (Image Credit: Cassini Imaging Team/SSI/JPL/ESA/NASA).}
\end{center}
\end{figure}
%
\subsection{The case of Icy Moons Enceladus and Europa}
\label{Encel_case}

While suspected since the early Eighties, when {\it Voyager} spacecrafts encountered Saturn's system, the existence of cryovolcanic activity 
on Enceladus has been revealed by {\it Cassini} spacecraft {\it Imaging Subsystem} \cite{porco_etal_2006}. The plumes, 
material ejected from Enceladus' South Pole (see Fig.~\ref{EncGeysers}), have been the subject of many works, and the analysis of their chemical 
composition showed a dominant role played by water vapor \cite{hansen_etal_2006,hansen_etal_2020}; other simple molecules like CO$_2$, 
NH$_3$, H$_2$ and CH$_4$ contribute for less than $\sim 10$\% 
\cite{waite_etal_2009,waite_etal_2017}. Recently, the data acquired by the mass spectrometers onboard the {\it Cassini} spacecraft, lead to the
detection of complex organic macromolecules, with masses above $200$~atomic mass units \cite{postberg_etal_2018}, like alkanes
bearing more than $14$ carbon atoms. These molecules have been found trapped in ice grains expelled by Enceladus' geysers. 
Whereas organic-free ice 
grains have a high salt concentration, those bearing organic material appear to be salt-poor grains. This latter type of grains does probably not 
reflect the composition of the Enceladus' salty ocean, in contrast with salt-rich grains generated by bursting of bubbles of volatile gas at water table. 
With molecular masses larger than $200$~amu, organic compounds cannot leave the liquid phase or be embedded in the gas flux 
before condensating from the vapor.
Alternatively, if they would be dissolved in the water, we should have detected their presence in all grains. 
Hence, the proposed and most plausible scenario \cite{postberg_etal_2018} is the existence of a thin film or floating organic layer 
on top of the water table. Over the surface of the Earth's oceans, the formation of this kind of microlayer is a well known process with a 
dedicated  literature (\textit{e.g.} \cite{marine_surface_films_2006}).

   The Enceladus subsurface water ocean is probably hidden below a $\sim 15$~km thick icy shell \cite{schubert_etal_2007}. 
At the interface between the liquid water and icy crust, the temperature should be around $273$~K, with an hydrostatic pressure  
reaching $1.7$~MPa. On its side, the gravity should not differ significantly from its surface value, {\it i.e.} $0.11$~m~s$^{-2}$.
Under these conditions the viscosity of liquid water remains around $1.6 \times 10^{-3}$~Pa~s \cite{pigott_2011}, because the influence
of pressure appears negligible \cite{pigott_2011}. Similarly, the surface tension of water is 
expected to be close to the room conditions value, {\it i.e.} $76$~mN~m$^{-1}$ \cite{kalova_mares_2018}.
   On the Earth, \rev{there are countless examples of hydrophobic natural organic matter}. For instance, in water treatment around $50$\% 
of organic compound present in raw water are hydrophobic \cite{kim_yu_2005}. Due to darwinian selection processes, many biological
surfaces are even superhydrophobic \cite{feng_jiang_2006}.

   Mechanisms at the origin of Enceladus organic compounds remain essentially unknown, but several hypotheses may be 
formulated. Carbonaceous chondrites are known to contain macromolecular organic compounds \cite{sephton_etal_2003,sephton_2004}, materials which have been
incorporated in the proto-Enceladus and altered afterwards by hydrothermal processes. Organic compounds may also be generated through abiogenic processes,
during fluid-rock interactions related to water circulation in Enceladus porous core \cite{choblet_etal_2017}. Two main geochemical routes are 
mentioned in the literature: the serpentinization and the Fischer-Tropsch type synthesis \cite{wang_etal_2014}. Both mechanisms may produce more 
or less complex hydrocarbons from a reservoir of CO and CO$_2$. Finally, the production of organic species by some sort of extraterrestrial form of life \cite{wagner_etal_2022} cannot be excluded.

Even if we are far to know the physical properties of Enceladus' organic material, some hints may be inferred from the properties of crude oils, that may be
considered as relevant analogs.These substances are complex multicomponent mixtures containing a large quantity of 
hydrocarbons. Among hydrocarbons, the size of alkanes ranges
roughly from C$_5$ to C$_{40}$ \cite{chilingar_etal_2005}, carbon chain lengths 
consistent with what we know about Enceladus. In general, raw petroleum is lighter than liquid water 
({\it e.g.} Tab.~5.2 p.~94, \cite{chilingar_etal_2005}), 
only very heavy oils reach a density larger than $0.9$, those above $1.0$ exist but are 
rather rare \cite{zou_2013}. Concerning their wettability, hydrocarbons show generally a high hydrophobicity
caused by the lack of polarity \cite{strom_etal_1987}.

Taken globally, the mentioned arguments draw the surface of Enceladus ocean as a 
very favorable place for organic material floatability, even under the form of solid particles possibly denser than water. Quantitative estimations can be also 
made, for instance about the maximum radius of floating particles, by applying Eq.~\ref{Rmax}. In a very unfavorable case, the density of the 
considered particle may be fixed to $\sim 1.2$ times the value of the water density. The contact angle of paraffin \cite{ray_bartell_1953},
namely $\sim 110^{\rm o}$, offers a representative value for solid hydrocarbons, 
depending on surface state of the sample. 
Applying Eq.~\ref{Rmax}, this angle leads to a maximum diameter of $12$~cm. Therefore, we can conclude safely that any organic particle
can float over the Enceladus water table surface, which appears as an ideal place for the formation of an organic floating film. This is a strong 
confirmation of the scenario proposed in the recent literature \cite{postberg_etal_2018}, as an explanation for the presence of organics in 
a fraction of icy grains of South pole plumes.

Among galilean Jupiter's moons, Io is probably the only one without a subsurface water ocean \cite{greenberg_2006}. In the remaining three
satellites, Europa has a particular place since its internal aqueous ocean \cite{vilella_etal_2020} is potentially connected to cryovolcanic 
activity \cite{roth_etal_2014,sparks_etal_2017}. 
In one decade, it will be explored by future space missions JUICE \cite{grasset_etal_2013} and Europa Clipper \cite{howell_pappalardo_2020}, 
and possibly complemented by a Europa lander mission currently under study.
According to thermal models, the temperature at the surface 
of Europa's ocean is close to $272$~K and the pressure ranges from $10$~MPa to $50$~MPa, depending on the assumed icy crust thickness
\cite{vilella_etal_2020}. Compared to Enceladus, water viscosity and surface tension should remain roughly unchanged since the pressure 
is not enough enhanced to lead to significant changes. The major difference is the gravity that is around $1.3$~m~s$^{-2}$, {\it i.e.} one 
order of magnitude larger than in the case of Enceladus. As a consequence, if some organic materials are also available 
at Europa water free surface; for a similar density and still according to Eq.~\ref{Rmax}, floating particles should have a maximum size $3$ times 
smaller than their Enceladean counterparts.

   The list of solar system objects discussed above, potentially harboring a liquid phase, is obviously not exhaustive, 
since ocean worlds seem to exist in the outer solar system \cite{nimmo_pappalardo_2016}. For instance, the Neptune's large satellite, Triton, has a surface 
where lava flows morphologic features have been identified \cite{croft_etal_1995}, while geysers expelling nitrogen have also been
observed \cite{soderblom_etal_1990}.
In this section, we have chosen to restrict our discussion to a few representative examples, supported by good observations and which could be 
explored in the near future.

\rev{
\section{Marine capillary waves}

    On a planetary science point of view, waves play an important role in ocean-atmosphere exchange of matter, heat, 
and momentum exchanges. Besides, they significantly contribute to costal erosion. Even though the question attracted the attention 
of scientists for centuries, the ocean waves generation remains a problem not fully understood in detail. 
This process is physically extremely complex: it involves an interface of two fluids of very different densities, the spatial scales 
vary from a few millimeters to kilometers, while time scales range from seconds to hours \cite{pizzo_etal_2021}. This high variability 
is mainly due to the presence of turbulent flows in both fluids. Not surprisingly, the waves pull their energy from wind and the growth 
process follows 3 steps: 1) wind turbulence applies random stress variations on the surface, these pressure and tangential shear 
fluctuations give rise to wavelets, due to resonances in the wind-sea coupling {\cite{phillips_1957,miles_1957}}; 
(2) the wave amplitude is then reinforced by the air flow {\cite{miles_1957}}; 
(3) the waves start to interact among themselves, exciting longer wavelength modes {\cite{komen_etal_1994}}. Recently, fully coupled direct 
numerical simulations were performed \cite{wu_etal_2022}. Capillarity is prominent during the early stages of wave formation, and when 
some kind of water atomization appears (wave breaking, whitecap). The second aspect will be addressed in the section~\ref{BubbleBursting},
since ``{\it bubble bursting}'' is recognized as a fundamental process for liquid water atomization. Here, we focus the discussion on
the onset of wave generation, partially driven by capillarity. The natural angular frequency of the capillary wave on a free surface 
is given by \cite{lamb_1993}
\begin{equation}
\label{omega0_cap}
    \omega_0^2 = \frac{\sigma k^3}{\rho_{\rm liq}}
\end{equation}
where $\sigma$ is the surface tension, $k$ is the wave number of the capillary wave, and $\rho_{\rm liq}$ is the liquid density.
It is difficult to go further without any of the properties of the wave ``excitator'', {\it i.e.}
the wind blowing over the sea surface. Nevertheless, Eq.~\ref{omega0_cap} draws clearly a difference between planetary bodies where 
the working liquid is methane and those harboring liquid water. The ratio $\sigma/\rho_{\rm liq}$ is found around $4 \times 10^{-5}$~m$^3$~s$^{-2}$
for liquid methane (relevant to Titan), for liquid water this value is about twice as large, {\it i.e.} $7\times 10^{-5}$~m$^3$~s$^{-2}$.
With these numbers we learn that, for a fixed wavelength, the wave growth should be facilitated by wind turbulences at different
natural frequencies.\\

  Due to the complexity of such air-liquid interactions, experimental approaches appear particularly relevant. Wind tunnel experiments, 
oriented to titanian and marsian planetary contexts were performed with the Mars Surface Wind Tunnel (MARSWIT) \cite{lorenz_etal_2005}. These
preliminary works suggested a strong dependence of capillary waves generation on atmospheric density, and waves on nonaqueous fluid surfaces looked 
more easily generated than on water. Another interesting discussion focussed on the Titan case, based on wind speed prediction provided by the 
TitanWRF Global Circulation Model and including the potential effects of viscosity and surface tension, 
was subsequently published \cite{lorenz_etal_2010}. Unfortunately, the wind speeds computed over Titan polar regions, together with the 
properties of sea liquid are still too uncertain to yield definite conclusion.
In the context of Cassini RADAR observations, a more sophisticated model of capillary-gravity waves on Titan's lakes has been 
proposed \cite{hayes_etal_2013}, leading to prediction of waves detectability as a function of seasons. Unfortunately, these predictions 
were not corroborated by RADAR measurements.

 For decades, researchers have been accumulating evidences or clues about the existence of very large areas of open water on 
the surface of Mars, during episodes of this planet geological history \cite{goudge_etal_2012,goudge_etal_2015}.
   The case of wave formation at the surface of a Mars primordial ocean is even more puzzling since the properties 
of atmosphere in ancient geological times, particularly its density, are not well known.
However, if Martian wave-cut shorelines could be confirmed, new constraints could be put on these properties \cite{banfield_etal_2015,adams_2023}.}

\section{Bubble bursting}
\label{BubbleBursting}
%
\begin{figure}[]
\begin{center}
\includegraphics[angle=0, width=8 cm]{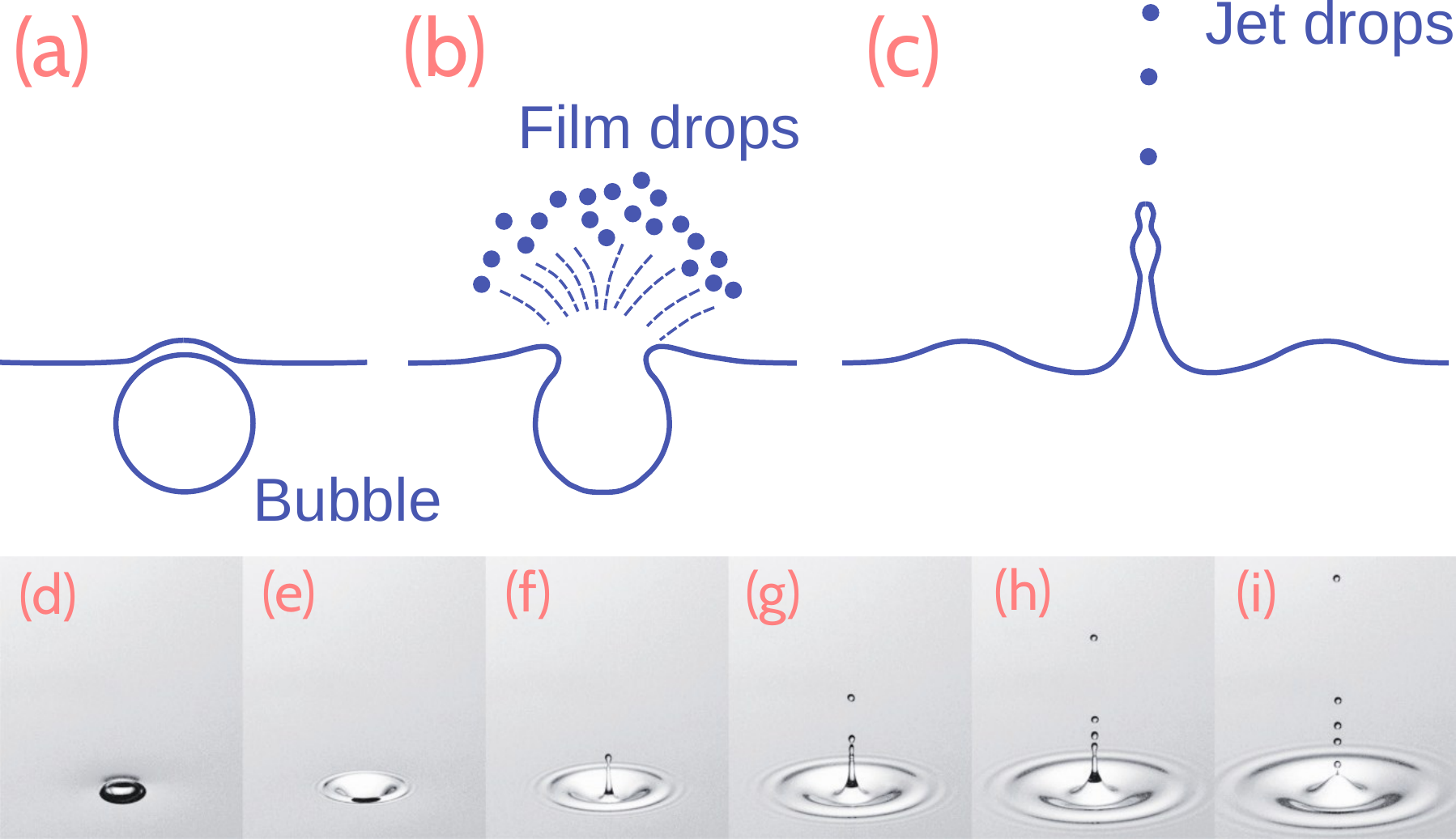}
\caption[]{\label{Fig3_Resch86_TS}(a)-(c) Main steps of drops production during bubble bursting (insprired from Fig.~3 in \cite{resch_etal_1986}).
           (d)-(i) Time-sequence of a millimetric floating bubble bursting at the surface of pure water and 
           propelling tiny jet drops above the surface (reprinted from \citeA{ref_3}).}
\end{center}
\end{figure}
%

Bubble bursting occurs in many natural and industrial processes, in physics, chemical and mechanical engineering, 
food science, geophysics, technology, medicine \cite{ref_1} and oceanography, field which is the most relevant here. 
On the Earth, bubble bursting is recognized to have an essential role in sea-spray formation, which process is a major source of
aerosols for the atmosphere at global scale and has a significant impact on 
the Earth’s radiative balance and biogeochemical cycling \cite{ref_2,ref_5,ref_6,wang_etal_2017}.
Since the pioneering study of bursting bubbles conducted by 
Woodcock et al. almost 70 years ago \cite{ref_2}, numerous experimental, numerical, and theoretical studies have been 
conducted with bubble bursting at a free surface. In brief, when a bubble reaches the free surface the residual liquid
film on the top part (Fig.~\ref{Fig3_Resch86_TS}-a) breaks ejecting a myriad of tiny droplets, nicknamed ``film drops'' (Fig.~\ref{Fig3_Resch86_TS}-b),
the cap collapse leading to the upward projection of the so-called Worthington jet, giving birth to larger ``jet drops'' (Fig.~\ref{Fig3_Resch86_TS}-c).
As a result of the well-known Rayleigh-Plateau instability, the liquid jet breaks up into a few droplets called jet drops, as seen in the 
time-sequence displayed in 
Figure~\ref{Fig3_Resch86_TS}-(d-i) \cite{ref_3}. \\
In recent decades, various scaling relations for the velocity, number and size of jet drops were reported in the literature. 
Nevertheless, it is only quite recently that much effort has been devoted to decipher minute details of the collapse of the immersed 
cavity leading to the production of jet drops during bubble bursting. Universal scaling laws relevant to the jet
dynamics and drop size produced by bursting bubbles were thus recently proposed through combined experimental and numerical approaches
\cite{ref_3,ref_22,ref_23,deike_etal_2018,ref_25}. 
These universal scaling laws were retrieved by using dimensional arguments based on the three following dimensionless numbers: 
The Weber number $We= \rho V^2 R /\sigma$, which compares the effect of inertia and capillarity, 
the Bond number $Bo= \rho g R^2 / \sigma$, which compares the effect of gravity and capillarity, 
and the Morton number $Mo= g \eta^4 / \rho \sigma^3$, which only depends on the fluid properties. 
In the previous dimensionless numbers, $V$ is either the jet velocity or the first jet drop velocity as it detaches from the tip of the jet, 
$R$ is either the mother bubble radius or the resulting jet drop radius, $g$ is the gravity acceleration, and $\mu$, $\rho$ and $\gamma$ 
are respectively the liquid dynamic viscosity, density, and surface tension. By using dimensional arguments and a collection of dedicated 
experiments, the jet dynamics was described through the following relationship, 
regime where the Morton number ranges over three decades ($10^{-11} \lesssim Mo \lesssim 10^{-8}$) \cite{ref_3,ref_23}:

\begin{equation}
  We_d \simeq 3.9 \times 10^{4} \times Mo^{2/7} \times Bo_{b}^{-1/2}
\end{equation}

with the subscript $b$ and $d$ referring to the bubble and drop, respectively. The latter equation yields the following dependence of 
the jet tip (and first drop) velocity $V_d$ with the mother bubble radius $R_b$ and liquid parameters (expressed in SI units):

\begin{equation}
  V_d \simeq 2 \times 10^2 \, \eta^{4/7} \, \sigma^{9/28} \, \rho^{-25/28} \, g^{-3/28} \, R_b^{-1}
\end{equation}

    In addition, in the same regime of Morton number, a scaling 
law was established between the Bond number of the first jet drop and a combination of both the Bond number of the bursting 
bubble and the Morton number as follows \cite{ref_22,ref_23}:

\begin{equation}
    Bo_d \simeq 1.1 \times 10^{-5} \, Mo^{-1/3} \, Bo_{b}^{6/5}
\end{equation}

     Finally, the latter equation yields to the following dependence of the first jet drop size as a function 
of the initial bubble radius and liquid parameters (expressed in the SI units):

\begin{equation}\label{r_jet}
  r_{\rm jet} \simeq 3.3 \times 10^{-3} \, \eta^{-2/3} \, \sigma^{2/5} \, \rho^{4/15} \, g^{-1/15} \, R_b^{6/5}
\end{equation}

   The minimum size $r_{d,{\rm min}}$ of jet drops produced by bubble bursting may be roughly estimated \cite{brasz_etal_2018} by considering 
the Laplace number $La= \rho \sigma R_b/\eta^2$ which takes the particular value $La_{\rm min} \simeq 7 \times 10^{3}$ for the 
minimum ratio \cite{brasz_etal_2018}, 
\begin{equation}\label{r_jet_min}
  \left(\frac{r_{\rm jet}}{R_b}\right)_{\rm min} \simeq 4 \times 10^{-3}
\end{equation}
allowing the derivation of $r_{jet,{\rm min}}$.\\

The production of film drops is a more complex mechanism not completely elucidated, even if fundamental works have 
been published on the topic \cite{bird_etal_2010}. Laboratory and marine field studies have smaller diameters roughly ranging from 
$1$~$\mu$m to $100$~$\mu$m and above \cite{resh_etal_1986}. While there are generally no more than $\sim 10$ jet drops
produced per bubble, the mean radius of film drops scales can be estimated using $\langle r\rangle_f \simeq R_b^{3/8} h_b^{5/8} /2$ and their 
number follows the law $N_f \simeq (R_b/l_c)^2 (R_b/h_b)^{7/8}$ with $h_b \simeq R_b^2/L$ where 
$L \simeq 20$~m on the Earth \cite{lhuissier_villermaux_2012}. The term $h_b$ represents the film thickness at the moment when 
the puncture occurs, the expression $h_b \simeq \mathcal{R}_b^2/L$ is explained by the convection within the film. Theoretical considerations
lead to $L \simeq l_c/\epsilon^{1/2}$ where $\epsilon$ is the efficiency of film puncture by convective cell (empirically $\epsilon \simeq 1.8 \times 10^{-8}$
for tapwater in laboratory conditions). The minimum size of film drops
can be evaluated using \cite{lhuissier_villermaux_2012}
\begin{equation}\label{f_fmin}
r_{f,{\rm min}} \simeq \left(324 \, \frac{l_c}{L} \right)^{1/3} \, l_c
\end{equation}
\\

Another important aspect of bubble bursting is its ability to eject materials dissolved 
in the liquid or present at the surface under the form of a more or less thick films. This subject has been explored by 
many works focused on the terrestrial case since it has a great importance in marine aerosol formation. Depending on the exact nature 
of the organic material, particularly if it is soluble or not in the liquid, the organic material might be detected in all aerosol particles 
formed by bubble bursting or not. The behavior of soluble particles 
has been studied in many works in the context of the Earth (\textit{e.g.} \citeA{bezdek_carlucci_1974,schmitt-kopplin_etal_2012,chingin_etal_2018}).
Even if no general quantitative law has been found, it has appeared that marine spray formation tends to enrich, with respect to the bulk 
chemical composition, droplets in surface-active organic matter when water is the solvant.\\
 The case of non-soluble compounds has been investigated in the context of oil spills 
\cite{ehrenhauser_etal_2014,liyana-arachchi_etal_2014}, for which 
the ejection to the atmosphere of organic material has been also attributed to bubble bursting. However, the mechanism is slightly
different because the alkanes are adsorbed at the liquid-air interface. Interestingly, the ejection of material trapped at this interface, via bubble 
bursting, seems to be independent of volatility and solubility of spill compounds.

\subsection{The case of Enceladus}

  As already recalled in Sect.~\ref{Encel_case}, Cassini instruments detected the presence of water ice grains in Enceladus' 
South pole plumes. The total mass of grains has been determined to be  $\sim 20$\% that of the water vapor \cite{dong_etal_2015}.
The sizes of these grains have been estimated with Cosmic Dust Analyzer (CDA) aboard Cassini spacecraft which has a detection threshold
around $0.2$~$\mu$m \cite{postberg_etal_2011}. With this instrument, a size distribution centered on a radius of $\sim 1$~$\mu$m has been
found \cite{postberg_etal_2018}. Further investigations have been carried out with the Cassini Plasma Spectrometer (CAPS) and the Radio and 
Plasma Wave Spectrometer (RPWS), the results have shown a grain size distribution peaked at $\sim 2$~nm \cite{dong_etal_2015}.

 These icy grains have been divided into three categories \cite{postberg_etal_2009,postberg_etal_2011} according to
their spectra: type I grains of almost pure water ice, type II grains containing a significant quantity of organic and/or siliceous 
material, and type III grains showing a very high salt concentration. Except for a portion of type I grains, which may be generated 
by homogeneous nucleation of icy crystals in the water vapor stream \cite{schmidt_etal_2008}, the formation of other grains demands 
an atomization of the Enceladus watertable surface, and a freezing during their ascent in geyser cracks \cite{postberg_etal_2009}.
 Such atomization may be obtained by two main processes: (1) a wave activity at the 
surface of liquid water, (2) bubble bursting of gases released by Enceladus core, or water vapor bubbles produced by surface boiling under 
low pressure.\\
    On the Earth, ocean waves are a major source of water aerosolization \cite{liu_etal_2021}. Such a wind-driven mechanism on Enceladus looks difficult to conceive. 
In contrast, \rev{processes} based on bubble bursting appear much more plausible. 
In Enceladus plumes, the mixing ratio of water vapor 
ranges from 96\% to 99\% \cite{waite_etal_2017} and represents an evidence for an evaporation process at work. Other species have been also detected 
in Enceladus' geysers by the Ion Neutral Mass Spectrometer (INMS) 
which was aboard Cassini, {\it i.e.} CO$_2$, CH$_4$, NH$_3$ and H$_2$, likely released by hydrothermal activity in the satellite core.
Thus, two categories of bubbles have to be distinguished: (1) bubbles containing the mentioned minor species originally formed in the core 
of Enceladus and traveling over tenth or hundredth of kilometers through the subsurface ocean, (2) the bubble of water vapor recognized to be 
generated within the few first tens of centimeters below the ocean surface by controlled boiling \cite{ingersoll_nakajima_2016}.
In the following, we discuss both cases, with a particular emphasis on subsequent sizes of droplets
produced by bubble bursting.

\subsubsection{Bursting of bubbles composed of trace gases}
\label{burstminor}

   Coalescence, pressure drop, or diffusive feeding by dissolved gases may increase the size of bubbles up 
to their break-up radius $R_{\rm bk}$ \cite{cordier_etal_2017a} during their ascent to the free surface. 
At Enceladus low gravity ($g_{\rm Enc}= 0.113$~m~s$^{-2}$), $R_{\rm bk}$ should be around $10$~cm, one order of magnitude larger
than the terrestrial value. 
The radius of subsequent jet drops produced by such big bubbles may be estimated using Eq.~\ref{r_jet}, yielding $r_{\rm jet, Enc} \sim 4$~cm. 
Even more striking is the corresponding minimum radius of jet drops given by Eq.~\ref{r_jet_min}, around $0.2$~mm.
Finally, Eq.~\ref{f_fmin} 
provides the minimum size of film droplets generated by Enceladus' mother-bubbles.
The capillary length for Enceladus is around $2.5$~cm, and if we assume that $\epsilon$, the efficiency of film puncture by convective cell, 
has a universal value and that the bubble size is around $10$~cm, the minimum film drop size should be around $8.8$~mm. As a consequence, 
streams of bubbles, with sizes comparable to the break-up radius, produce liquid aerosols by bursting orders of magnitude too large to explain 
the observed sizes of Enceladus' geysers ice grains ($\sim 1 \mu$m).
 The bubbles initially formed in Enceladus' core, likely by heterogeneous nucleation, may have relatively
small sizes, and their number per unit of volume of the subsurface ocean may be sufficiently low to impede significant coalescence,
in such a way that bubbles reaching the free surface are tiny enough.
However these circumstances remain very hypothetical, and do not impede other mentioned growth processes.
Then, we have to review physical effects that could lead to small bubble bursting.\\
     First, during their ascent to the free surface, bubbles undergoing a break-up event, produce an 
undetermined number of smaller bubbles. Several effects can potentially induce such break-up, for instance those driven by 
hydrodynamics stresses or surface instabilities \cite{chu_etal_2019}, but the corresponding size distribution cannot be evaluated.
     Second, bubble-bursting cascades may occur at sea surface. As show by laboratory experiments \cite{bird_etal_2010}, 
following the rupture of the initial bubble, a ring of smaller bubbles can appear and the phenomenon may repeat, giving birth to a second 
generation of tiny bubbles. This two-step cascade seems to be able to reduce the diameter of bursting bubbles by two orders of magnitude 
\cite{bird_etal_2010}, a diameter decrease which is not necessarily sufficient to get micro- or nano-metric ejected droplets.\\
   Recently, a new mechanism based on the flapping shear instability has been proposed to explain the submicron drops in the 
context of the Earth oceans \cite{jiang_etal_2022}. This mechanism seems to be efficient in producing film drops down to a few nanometers.
Unfortunately, this process depends on the density ratio $\rho_{\rm air}/\rho_{\rm liq}$ which has to be larger than $\sim 10^{-3}$. This
condition is fulfilled in the terrestrial context with an air density around $1.2$~kg~m$^{-3}$, but on Enceladus, with a water 
vapor ``atmosphere'' it looks impossible. Right above the sea surface, the pressure can be at most equal to the water vapor pressure which is $\sim 611$~Pa 
at $0^{\rm o}$C, corresponding to a density of $4.9 \times 10^{-3}$~kg~m$^{-3}$, well below the terrestrial air value. The required criterion is therefore
not satisfied by around $3$ orders of magnitude.\\
     In sum, according to the general picture we have drawn about trace gas bubbles, we have to investigate other possibilities
which could lead more confidently to the production of micrometric droplets.

\subsubsection{Bursting of water vapor bubbles}

     Concomitantly, the low pressure of the vapor phase right above the ocean causes the boiling of the very first layers of
liquid, typical over a few tens of centimeters \cite{ingersoll_nakajima_2016}. Unfortunately, we found very few references bringing information 
about the size distribution of these water vapor bubbles. In some laboratory works, related to nuclear power plant systems, 
the size of nucleating vapor bubbles seems to be around $0.5$~mm 
(\textit{e.g.} \citeA{kazuhiro_etal_2017}) but the conditions are different
compared to those of Enceladus ocean, particularly concerning the nature of nucleation sites. On Enceladus these sites should composed by water ice
or organic material while in terrestrial experiments the solid substrates are made of steel or glass. This point is important since the 
size of bubbles leaving the nucleation substrate is related to the contact angle which depends on the chemical nature of the material
(see for instance \citeA{giraud_2015}). 
It can be shown that, for a given horizontal solid surface, the detachment radius, {\it i.e.} the value 
reached by the bubble when it leaves its substrate, is given by
\begin{equation}\label{myRdetach}
    R_{\rm detach} \simeq \sqrt{\frac{3 \, \sigma}{2 \, \rho_{\rm liq} g}} \, \sin \theta_{c}
\end{equation}
with $\sigma$ the surface tension and $\theta_{c}$ the contact angle. This equation is very similar to those proposed in 
the literature \cite{giraud_2015,fritz_1935}.
The contact angle between water ice and liquid water appears to be very close to zero \cite{ketcham_Hobbs_1969,knight_1971}, leading to 
extremely small values for $R_{\rm detach}$. If we assume, for instance, a contact angle of $\sim 1^{\rm o}$, Eq.~\ref{myRdetach} leads to
$R_{\rm detach} \simeq 5.4 \times 10^{-4}$~m. 
The minimum size $r_{jet, min}$ of jet drops may be approached using Eq.~\ref{r_jet_min}, this yields $r_{\rm jet, min} \sim 2$~$\mu$m. Clearly, 
for a contact angle below $0.5^{\rm o}$, jets generate submicron drops. The vertical velocity of such tiny bubble, in their ascent to the 
free surface, is very low and more than $\sim 4$~years can be required for a micrometer-sized bubble \cite{fifer_etal_2022} to cover a distance 
of $\sim 71$~cm \cite{ingersoll_nakajima_2016} through a quiet pool.\\
   In fact, the first subsurface layers should be vigorously mixed by macroscopic convective eddies. We have identified several phenomena 
inducing mixing: The ascent of large bubbles coming from depths of Enceladus, which may drag the liquid and small water vapor bubbles, 
and the travel of centimetric methane bubbles through the last $\sim 70$~cm is around $\sim 10$~seconds. 
In addition, to maintain 
an evaporation flux, heat has to be transported from the ocean bulk to its surface. Otherwise, due to evaporative cooling, the surface 
freezes, stopping the geyser activity. Such a heat flux implies macroscopic convection. Finally, after evaporation, the remaining liquid water
is cooler and more salty, therefore denser, and it should sink. Besides, during its journey to the surface, a tiny water vapor bubble should grow.
The resulting size of water vapor bubbles bursting at the surface is the fruit of a competition between their growth and transport to the 
surface. Making an estimation of a relevant time scale is not easy. Nevertheless, the bubbles nucleating 
close enough to the surface will have a limited growth, and 
convection is characterized by eddies with non-uniform velocity distributions,
letting the fastest eddies transport small bubbles with restricted growth. In conclusion, it is very plausible that bubbles of water 
vapor, small enough to produce micrometer-sized water droplets corresponding to film drop bursting, reach the water free surface within 
the Enceladus cracks, these liquid aerosols producing subsequently micrometer-sized icy grains.\\ 
    Besides, organic materials may also be present at the surface of the Enceladus ocean, or in its immediate subsurface vicinity. Water vapor bubbles
may also nucleate at these organic surfaces. In this case, the expected contact angle should be much larger than in the case of water ice, 
result which is particularly true when the organic matter is hydrophilic. For instance, with $\theta_{c} \sim 90^{\rm o}$, $R_{\rm detach}$ reaches
$3$~cm, corresponding to a minimum jet drops size above $\sim 100$~$\mu$m.\\
   In closing, water vapor bubbling appears to be the most favorable process to produce
water droplets small enough to account for the existence of micrometer-sized icy grains in the Enceladus geyser plumes, particularly in those
containing salts.

\subsubsection{Organic matter aerosolization}

\begin{figure}[]
\begin{center}
\includegraphics[angle=0, width=8 cm]{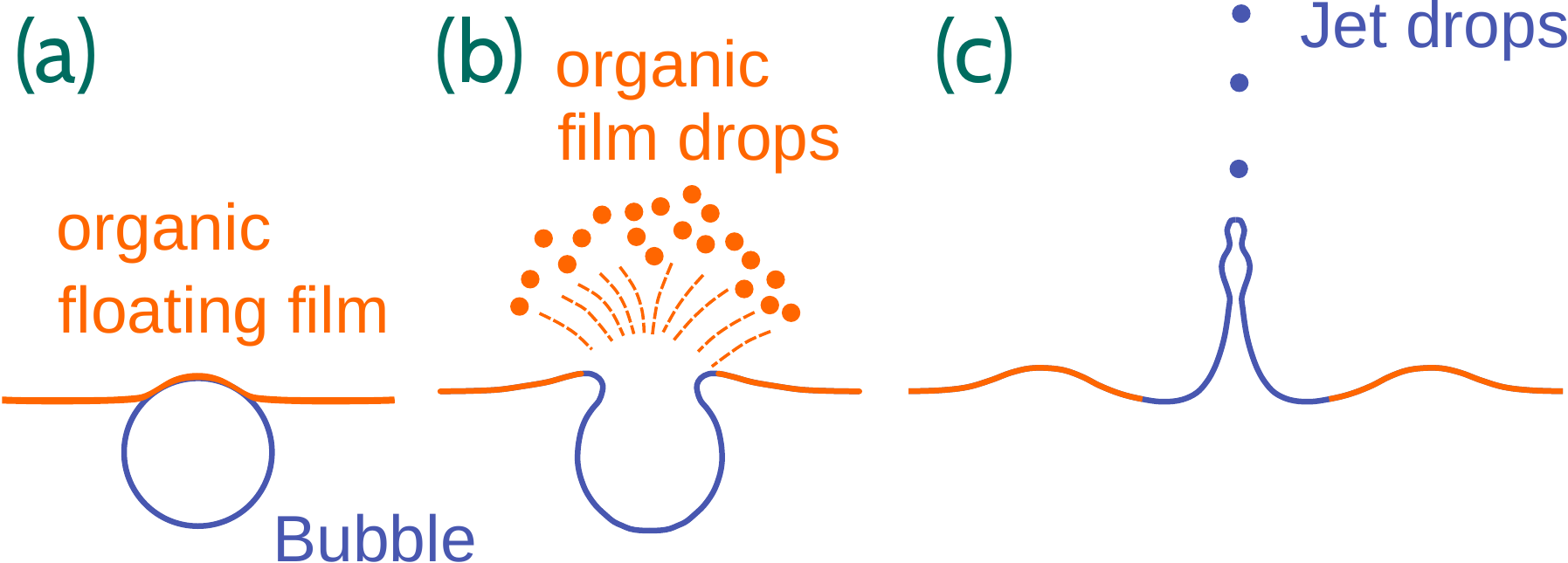}
\caption[]{\label{FilmDropsOrga}A conceptual illustration of bubble bursting in the presence of a surface floating film, this 
          sketch follows essentially Fig.~\ref{Fig3_Resch86_TS} (a)-(c). Cases (a) and (b) correspond to a situation where
          the thcikness of the organic film is small compared to the bubble diameter, even if this film is not necessarily
          a monomolecular deposit. 
          Here, only tiny film droplets of pure organic material are produced but, if the thickness of slick were larger than the 
          bubble dimensions, jet drops would be also ejected.}
\end{center}
\end{figure}

   As already recalled in Sec.~\ref{Encel_case}, organic materials probably colonize, at least partially, the water surface in
Enceladus cracks. These organics could have been transported thanks to their own buoyancy, or drained from the moon deep interior by 
trace gases bubbles. 
The organic material could be molecules or small solid particles. In the first case, molecular surfactants can easily 
cover the surface of ascending bubbles; in the second case, a surface adsorption, driven by capillary forces as described in Sec.~\ref{floata_capill}, 
can be invoked. As a consequence, organic deposits of various thickness could be built up at the water free surface, ranging from (1) organic layers thicker
than bubble dimensions and (2) spills with a thickness large compared to molecular sizes but relatively small in comparison to the bubble diameters, 
to (3) molecular carpets. We can wonder how bubbles of water vapor, or containing trace gases, interact with these floating materials.\\
   In the first case, depending on the physical state of the material (liquid or solid) and on the size of continuous pieces of matter, 
the interaction with bubbles will have a different character. The bubbles will not be able to go through a solid ``organic pack'', while their 
penetration in a floating organic liquid can be considered. The latter scenario is only valid for species still liquid at $\sim 273$~K, which 
should have a relatively small molecular weight. For instance, in the case of linear alkanes, the temperature of 
fusion (See the database \url{https://webbook.nist.gov}) remains below $\sim 273$~K for molecules with a chain shorter than 
C$_{13}$ (tridecane). If the floating film has an horizontal extension much larger than the typical size of bubbles, 
then bubble-bursting may produce droplets of pure organic material (see Fig.~\ref{FilmDropsOrga}). 
In the situation hypothetized in Fig.~\ref{FilmDropsOrga} sketch, laboratory
experiments has brought to light the formation of a thin oily thread associated with jets during bursting in a system containing water covered
by an oil film \cite{ji_etal_2021}. This effect could generate small mesoscopic organic aerosols in the Enceladus context.
In the case where the slick has a thickness larger than the typical size of bubbles, ``pure'' jet drops 
of organic may be produced. According to Eq.~\ref{r_jet}, it is not obvious how the size of organic material jet drops could compare to their aqueous analogs since 
the organic liquid may be less viscous and less dense. However, the previous discussion about sizes of jet and film drops 
still applies to organic liquids, opening the possibility for the formation of small organic aerosols. If small enough, aerosol particles can be embedded 
in the vapor streams ascending along the cracks. During their journey, these drops can solidify and form small solid particles which could serve as
water ice freezing nuclei. A similar reasoning may apply to floating solid phases: small enough 
solid particles may be embedded in geyser ascending streams, sticked to water droplets. We emphasized that mass spectra of organic materials
have been found by Cassini instruments up to $1000-2000$~u, and, due to the grain fragmentation in the instrument, the parent particles should 
have contained even many more larger particles \cite{postberg_etal_2018}. This scenario is consistent with the existence of small solids.\\
  A very interesting aspect has been revealed a decade ago by laboratory experiments: in the presence of surfactants in the water.
An organic deposit at the liquid-air interface can lead to the production of nanoemulsions in the surface layers of the ocean
\cite{feng_etal_2014}. These nanoemulsions could take the form of submicrometer-sized organic droplets dispersed in the water, material that 
could be subsequently ejected into the air by bursting. The presence of the required surfactants has been suggested by plume
measurements \cite{postberg_etal_2018}\\
     Up to this point we did not consider the possible chemical fractionation between organic material at the surface of the ocean,
and organic compounds aerosolized. We emphasized that this question is significantly different 
from that studied in recent investigations \cite{fifer_etal_2022}, in which the relantionship between plume 
composition and ocean composition is explored by ignoring bubble-bursting mechanisms. 
The possible chemical fractionation during bubble-bursting is still a matter
of scientific debates. However some general trends may be drawn. For instance, concerning relatively thick slick, the fractionation should
be a minor effect. This behavior is suggested by observations of oil spills on the Earth.
In such event, bubbles are produced by breaking waves and it has been shown that the aerosolization of organic matter is pratically independent of the vapor 
pressure and chain length of organic components \cite{ehrenhauser_etal_2014,liyana-arachchi_etal_2014}. 
Excluding exceptional circumstances like oil spills, the biological activity enriches the sea surface layers with a plethora 
of organic species, and question of their transfer to aerosols has been largely studied, since this process has a major 
role in the physico-chemistry of the atmosphere, and influences the climate (\textit{e.g.} \citeA{bezdek_carlucci_1974,ganan-calvo_2022}).
It has been shown that surface active molecules tend to be overconcentrated in liquid 
aerosols compared to bulk water \cite{bezdek_carlucci_1974}. This seems to be a general result. For instance it has also been found in
studies dedicated to the effervescence of champagnes \cite{ligerbelair_etal_2009}. In addition, a chemical fractionation appears to be at work 
\cite{schmitt-kopplin_etal_2012,chingin_etal_2018,crocker_etal_2022}. Such mechanisms could also be present on Enceladus, but it is difficult to
disentangle what could be associated with mesoscopic or macrosopic, grain of organics, or with molecular floating films.
Perhaps, far future space missions like 
{\it Exobiology Extant Life Surveyor} 
may bring clues on the topic. Nonetheless, bubble bursting is also a very plausible scenario for the existence of the Enceladus type II (organics rich) 
grains.\\
%
%

\subsection{The case of Titan}
\label{flapping_Titan}

   In the case of the Earth, the ocean surface agitation, through several atomization mechanisms, dominates the sea/atmosphere 
water exchange when the wind speed exceeds $\sim 15$~m~s$^{-1}$ \cite{andreas_etal_1995}. It is difficult to imagine such conditions
over Titan's polar seas 
where marine surfaces are rather calm with an apparent absence of waves
\cite{wye_etal_2009,zebker_etal_2014,grima_etal_2017}. However, even if bubble bursting is not induced by breaking waves, streams
of nitrogen rich bubbles may come into play \cite{hofgartner_etal_2014,hofgartner_etal_2016,cordier_etal_2017a}. These bubbles 
could interact with postulated floating layers \cite{cordier_carrasco_2019}, and propel material into the atmosphere. A similar process
is at work on the Earth with marine biofilms which are sources
of organic aerosols. Moreover, similarly to the 
Enceladus case, the relatively low gravity of Titan ($1.352$~m~s$^{-2}$) should promote large bubbles, giving birth to large jet drops.\\
   However, very noticeable differences between Titan and Enceladus or the Earth, are the nature of the liquid 
( liquid methane instead of water) and the density of the atmosphere in contact with the ocean.
As already mentioned in Sec.~\ref{burstminor}, the flapping shear instability \cite{jiang_etal_2022}, producing submicron 
film drops, in terrestrial environment, is expected to arise on Titan. Indeed, in this case 
$\rho_{\rm air} \simeq 5$~kg~m$^{-3}$ and $\rho_{\rm liq} \simeq 500$~kg~m$^{-3}$, leading to the ratio $\rho_{\rm air}/\rho_{\rm liq} \simeq 10^{-2}$,
one order of magnitude above the critical value of $\simeq 10^{-3}$ 
(see Fig.~3-A in \citeA{jiang_etal_2022}).
This criterion has to be complemented 
by a second one:
the radius of the mother bubble must be smaller than a fraction of the local capillary length ($\simeq 8$~mm on Titan).
In the case of Titan bubbles have to be subcentimetric, which is not a very restrictive condition. Indeed, the break-up
diameter for bubble rising in Titan's seas is of the order of $\simeq 4.6$~cm \cite{cordier_etal_2017a}, which has to be considered as a maximum value.
Statistically it leaves space for populations of centimetric or subcentimetric bubbles.
The theory of the ``flapping shear instability'' leads, for an initial gas bubble with a radius of $5$~mm, to an average diameter of 
produced droplets around $30$~$\mu$m.
With a gravity and a liquid density lower than their terrestrial counterparts, 
the minimum vertical wind velocity to uplift a micrometric drop in the titanian atmosphere is around a few millimeters per second. 
This aspect opens the door to a potential redistribution of Titan's sea material at global scale extension. Any kind of sea molecule 
could be transported to any point of the titanian surface and be involved in microphysical processes of the atmosphere. 
Finally, this ocean-atmosphere interaction mechanism offers the possibility for global 
surface seeding by molecules generated/processed in the deep interior of Titan, and brought to its seas by geophysical phenomena like icy 
crust convection \cite{kalousova_sotin_2020}. It can not be excluded that such molecules could be detected 
in regions explored by the {\it Dragonfly} mission.

\section{Extraterrestrial rain droplets}

   The existence of an hydrological cycle has a huge impact on landscape morphology of planetary bodies. 
In the solar system only our planet, and one moon, Titan, are harboring such an active cycle at present. 
Besides, Mars exploration has revealed numerous evidences of a past hydrological activity.
Indeed, ancient terrains of Mars preserve landscapes consistent with valley networks, lakes, alluvial fans, deltas 
\cite{fassett_head_2008,howard_etal_2005,irwin_etal_2011,hynek_etal_2010} and possibly even oceans \cite{parker_etal_1993}. These 
morphological features support the presence of liquid water flowing on the Martian surface during Late Noachian to Early Hesperian periods, 
$\sim 3.6$ to $4$~billion years ago. The widespread occurrence of clays is also a strong evidence for persistent water on ancient Mars \cite{bibring_etal_2006,carter_etal_2015}.
A warmer and wetter climate than today, at least episodically, is required to explain these geologic evidences, with an active hydrological cycle.
Nowadays, liquid water may exist in the depths of Mars \cite{lauro_etal_2021} or intermittently at its surface 
\cite{martintorres_etal_2015,ojha_etal_2015}. However, the current amounts involved are, by far, too small to sustain a hydrological cycle.\\
  Generally speaking, and leaving aside the particular case of hail, rainfalls are the cycle phase during which water returns 
to the planetary surface. Due to the importance of weather predictions, the properties of rain droplets have caught the interest of 
scientists for a long time \rev{\cite{bentley_1904,marshall_palmer_1948,loftus_worsworth_2021}}. Since the interaction between raindrops and the soil 
depends on drop properties, we focus our attention on these properties.

\subsection{Size distribution of rain droplets}
\label{sizedistrib}

   Probably the most salient and basic features of raindrops are their average size together with their size 
distribution. Early works focused on their experimental determinations \cite{laws_parsons_1943,marshall_palmer_1948}, lead to 
a number density of drops $n(d)$ (cm$^{-4}$), corresponding to sizes between $d$ (counted as diameter) and $d + \mathrm{d}d$ per unit volume 
given by \cite{villermaux_bossa_2009}
\begin{equation}
       n(d) = n_0 \, \exp (-d/\langle d\rangle)
\end{equation}
with $n_0$ (cm$^{-4}$) a quantity reflecting the mean spatial density of drops, and $\langle d\rangle$ (cm) the average drop diameter.
This distribution law compares nicely with empirical data for drop diameters $d \gtrsim 1$~mm.
On the Earth, at ground level, the constant $n_0$ has a value around $0.08$~cm$^{-4}$ \cite{villermaux_bossa_2009}, but it varies 
with temperature \cite{houze_etal_1979}. The average diameter $\langle d\rangle$ is empirically related to the rainfall precipitation 
rate $\mathcal{R}$ (mm~h$^{-1}$) by
\begin{equation}\label{d_bar}
    \langle d\rangle^{-1} = 41 \, \mathcal{R}^{-0.21}
\end{equation}
These laws have been confirmed by in situ measurements \cite{mason_1971,houze_etal_1979,ulbrich_1986} employing various methods, and is a 
broadly accepted result. The mean size of drops depends on the precipitation intensity: on average a 
heavy storm produces larger hydrometeors than a fine drizzle. Very interestingly, a quantitative explanation of prefactor and exponent 
values in Eq.~\ref{d_bar}, based on first principles, is available in the literature \cite{villermaux_bossa_2009}. Then, the 
developed formalism makes predictions possible for extraterrestrial contexts.
  In this frame, the precipitation rate $\mathcal{R}$ may be expressed as 
\begin{equation}\label{Rtheo}
    \mathcal{R} = n_0 \frac{\pi}{6} \, \sqrt{\frac{\rho_{\rm liq}}{\rho_{\rm air}}} \sqrt{g} \, \langle d\rangle^{9/2} \,
                  \int_0^{+\infty} x^{7/2} p(x) \, \mathrm{d}x
\end{equation}
where the integral represents the shape of the drop size distribution and has an approximate value of $28.34$
\cite{villermaux_bossa_2009}. Other quantities are $\rho_{\rm liq}$ and $\rho_{\rm air}$, respectively the liquid and the 
air densities (kg~m$^{-3}$), and $g$ the gravity (m~s$^{-2}$). Parameters specific to the Earth yield to
\cite{villermaux_bossa_2009}
\begin{equation} 
    \langle d\rangle^{-1} \simeq 48.5 \, \mathcal{R}^{-2/9}
\end{equation}
strikingly close to the empirical relation provided by Eq.~\ref{d_bar} ($2/9 \simeq 0.22$). If the fragmentation processes drive the size distribution
of raindrop, the integral in Eq.~\ref{Rtheo} with $x = d/\langle d\rangle$ and 
\begin{equation} 
   p(x) = \frac{32}{3} \, x^{3/2} \, K_3(4\sqrt{x})
\end{equation}
should be universal and, consequently, it should not depend on the planetary context. Nevertheless, the pre-factor 
\begin{equation} 
   f = \sqrt{\frac{\rho_{\rm liq}}{\rho_{\rm air}}} \sqrt{g}
\end{equation}
depends on the planetary context. For our planet, we have the value $f_{\rm Earth} \sim 90.4$~m$^{1/2}$~s$^{-1}$, while for Titan 
$f_{\rm Titan} \simeq 10.8$~m$^{1/2}$~s$^{-1}$ (for numerical inputs, see \citeA{cordier_etal_2017a}). Thus, for a fixed precipitation rate $\mathcal{R}$, 
the value of $f$ factor suggests mean raindrop diameter $\langle d\rangle$ $\simeq 1.6$ times larger in Titan's rainfalls. Another quantity, 
relevant for raindrops features, is the maximal size $d_{\rm max}$ of a stable drop given by \cite{villermaux_bossa_2009}
\begin{equation}\label{dmax}
   d_{\rm max} = \sqrt{\frac{6 \sigma}{\rho_{\rm liq} g}}
\end{equation}
yielding a diameter of $6$~mm in the terrestrial case, while for Titan we get $d_{\rm max} \simeq 1.3$~cm. This evaluation, not dependent on the 
precipitation rate, reinforces the picture of larger raindrops on Titan. This result is consistent with early estimations 
\cite{lorenz_1993} based on similar principles.\\
The case of early Mars is more tricky since 
the thermodynamic conditions (pressure and temperature in its atmosphere of CO$_2$-H$_2$O), 
that prevailed in this epoch are not well established \cite{ramirez_etal_2014}. Nitrogen and argon isotope 
compositions, measured in the Martian meteorite Allan Hills 84001, provide a lower limit of $0.5$~bar for the atmospheric ground pressure 
$4$~billion years ago on Mars \cite{kurokawa_etal_2018}. Of course, the existence of liquid water requires a temperature around $273$~K.
The value of the ground pressure is more uncertain, but 
a range between $0.5$ and $4$~bar appears to reasonably bracket the real values \cite{ramirez_etal_2014,craddock_lorenz_2017}.
Relevant numerical values yield
a ratio $f_{\rm Mars}/f_{\rm Earth}$ in between $0.3$ and $0.7$.
As a consequence, for an equivalent precipitation rate $\mathcal{R}$, the average droplet diameter should be only $\sim 20$\% larger in the early Mars 
context than on our nowadays planet. Concerning the raindrop maximal size, we found for Mars $d_{\rm max} \simeq 1$~cm, roughly 
twice as large as the diameter of 
terrestrial drops. This estimation is consistent with previous determinations \cite{craddock_lorenz_2017}, particularly if we take into account 
the statistical significance of such an assessment.\\
     Here, both estimations, for Titan and Mars, have been made assuming a similar parameter $n_0$. The parameter $n_0$ and precipitation rate $\mathcal{R}$ depend
on the details of precipitation microphysics, and more generally on the global behavior of the atmosphere.
     In the following, we examine several possible geophysical consequences of large raindrops.
     
\subsection{Raindrop-Induced Erosion}
\label{rainInducEros}

   \rev{With tectonic activity and volcanism, erosion is one of the major geophysical processes that shape a planetary landscape. 
Erosion cannot be reduced to a single phenomenon, it can be governed by chemical dissolution or induced by mechanical effects \cite{lorenz_lunine_1996}.}
On the Earth, erosion is caused by many phenomena combining the effects of winds, precipitation or runoff of liquid water, and the effects of water ice during
freezing or melting. Raindrop splash can be seen as the first step in erosion of soil by rainfalls. 
Unfortunately, even the erosion provoked only by rainfalls depends on endless parameters: soil erodibility, slope steepness,
rainfall rate, etc \cite{caracciolo_etal_2012,gholizadeh_etal_2018,wang_etal_2020}. All these parameters, empirically determined in the terrestrial case, are 
absolutely unknown, even when they could be relevant, in extraterrestrial environments. The only aspect we can discuss here is the raindrop 
kinetic energy which is a parameter involved in raindrop-induced erosion, since it characterizes the strength of the impacts, and the ability of a
droplet to detach particles from soil.
For ease of comparison, we have adopted $d_{\rm max}$ (see Eq.~\ref{dmax}) as the size of a droplet and the velocity of falling is provided by the terminal velocity
\cite{lorenz_1993}
\begin{equation}
 U = \sqrt{\frac{2 m_{\rm drop} g}{\rho_{air} S C_{d}}}
\end{equation}
with the mass of the raindrop $m_{\rm drop}= \pi \rho_{\rm liq} d^3/6$, the drag reference area $S$ and the drag coefficient $C_{d}$ 
\cite{clift_etal_1978}. The deformation due to aerodynamic forces has been taken into account for the estimations of both $S$ and $C_{d}$ 
\cite{lorenz_1993}. \rev{A slightly different drag parametrization may be found in more recent works \cite{loftus_worsworth_2021}.}

In the case of the Earth, we found $U_{\rm Earth} \simeq 9$~m~s$^{-1}$, and a terminal kinetic energy $K_{\rm Earth} \simeq 4.5$~mJ.
For Titan, the relevant numerical inputs lead to $U_{\rm Titan} \simeq 1.5$~m~s$^{-1}$ and $K_{\rm Titan} \simeq 0.65$~mJ.
Finally, in the Marsian context we have considered two values of pressure ({\it i.e.} $0.5$ and $4$~bar, see Sec.~\ref{sizedistrib}) yielding respectively 
$U_{\rm Mars} \simeq 7.8 - 2.6$~m~s$^{-1}$ and $K_{\rm Mars} \simeq 10^{-2} - 1.1$~mJ.
All other parameters remaining equal, the maximum kinetic energy for raindrops seems to be significantly lower in the context 
of Titan than in that of Mars or the Earth. However, it remains much larger than the erosivity threshold, 
{\it i.e.} the minimum kinetic energy (around a few $\mu$J)
required to initiate soil particle detachment as experimentally determined for fine sand and silt loam \cite{salles_etal_2000}. 
These estimations suggest possible weaker raindrop-induced erosion on Titan, compared to the Earth 
or Mars, but it is impossible to be more conclusive in the present state of our knowledge. Titanian erosivity 
thresholds for relevant organic analogs should be measured in laboratory; but, river bed networks observed on Titan indicate that significant 
erosion by runoff has been at work in the past.

We emphasize that the kinetic energy mentioned here is
a drop individual quantity and does not include, the precipitation rate as it was done in other estimations \cite{whitford_duval_2020}.

\subsection{Raindrop imprints}
%
\begin{figure}[]
\begin{center}
\includegraphics[angle=0, width=9 cm]{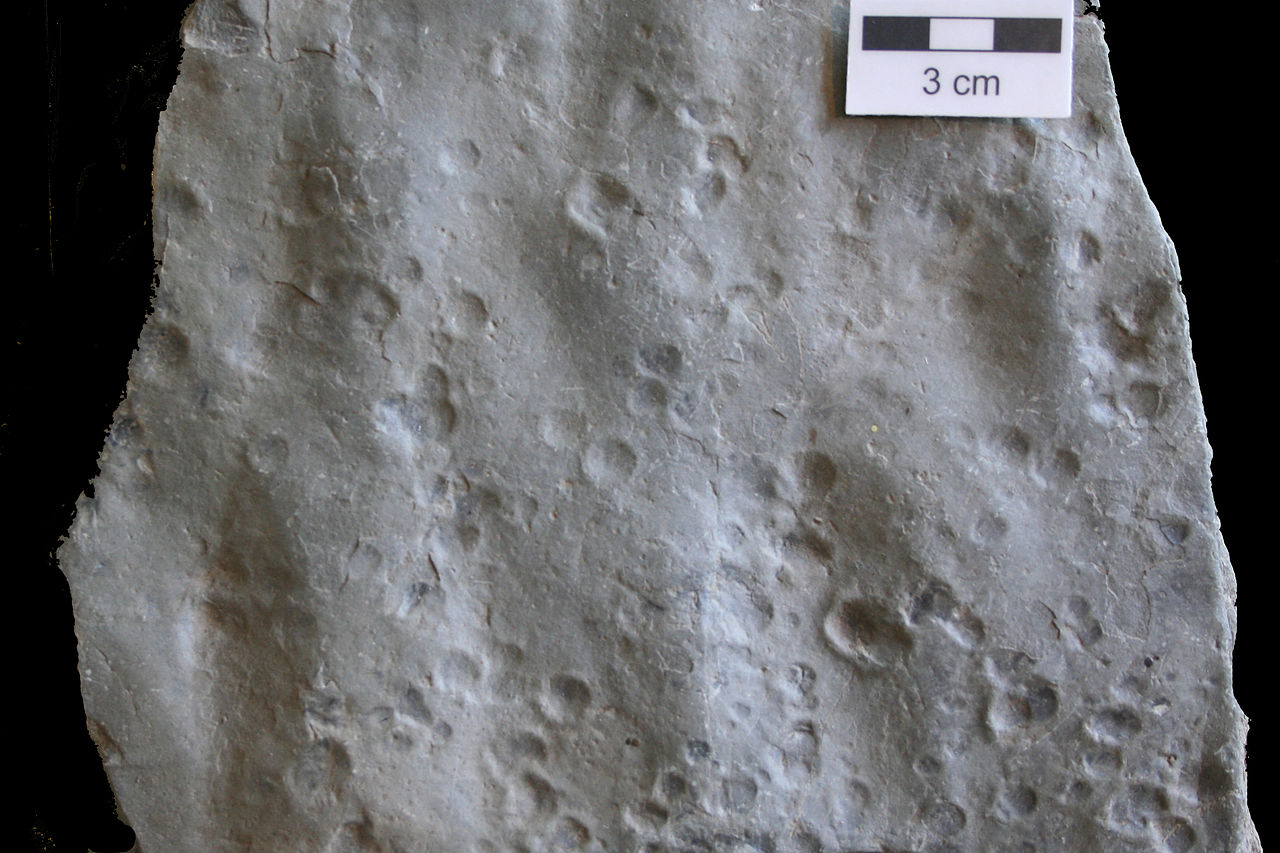}
\caption[]{\label{RainDropImpre}An example of fossil raindrop imprints on the top of a wave-rippled sandstone 
           from the Horton Bluff Formation (Mississippian). This sample is on display at the Blue Beach Fossil Museum (Hantsport, Nova Scotia).
           Picture taken by Michael Rygel, publicly shared under Creative Commons Attribution-Share Alike 3.0 Unported license.}
\end{center}
\end{figure}
\begin{figure}[]
\begin{center}
\includegraphics[angle=0, width=14 cm]{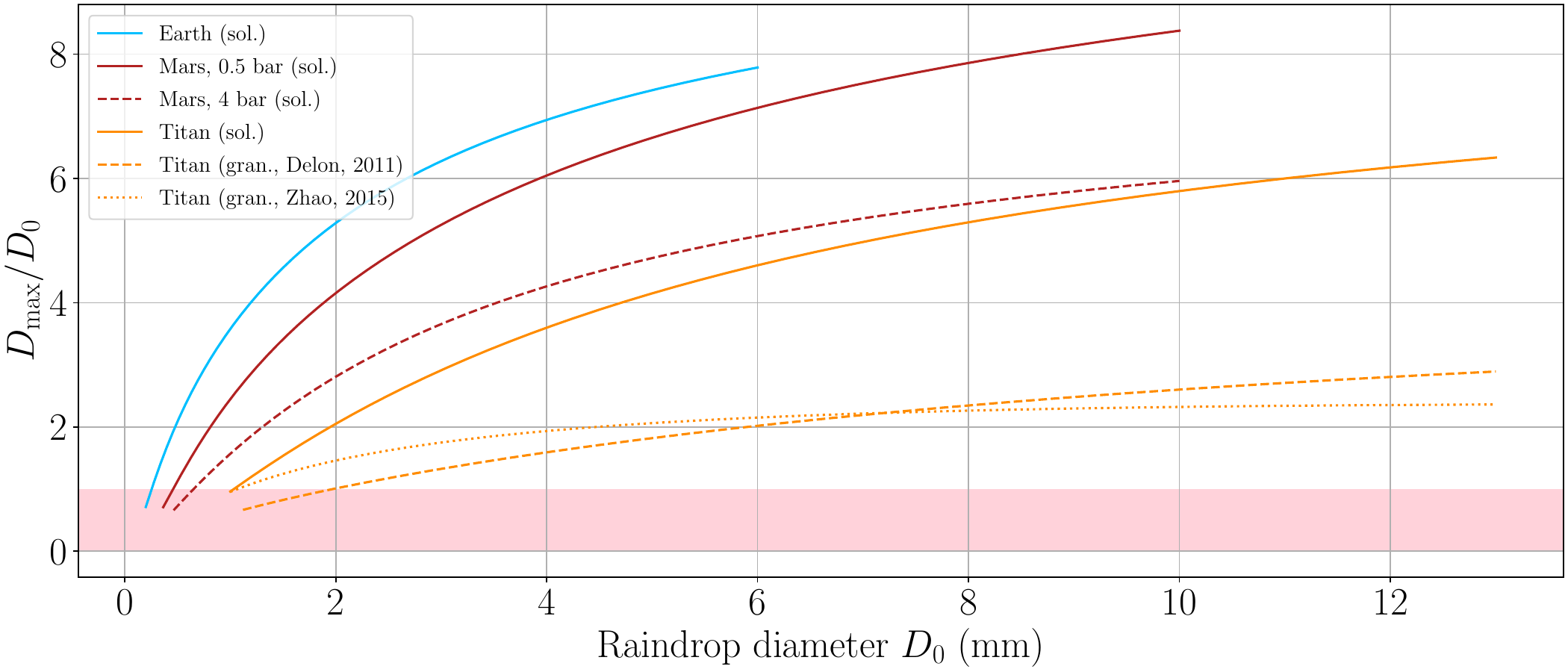}
\caption[]{\label{FigDmaxD0}The ratio $D_{\rm max}/D_0$, with $D_{\rm max}$ the maximum diameter of a raindrop of initial 
           diameter $D_0$, spreading when impacting a solid surface (``sol.'' label). Two atmospheric pressures have been considered in the case of Mars.
           We have also plotted curves showing the variations of $D_{\rm max}/D_0$ when a raindrop impacts
           a granular medium (``gran.'' label), the corresponding prescriptions have been found in \citeA{delon_etal_2011} and \citeA{zhao_etal_2015}.
           For the latter a ``packing fraction'' of $0.58$ has been hypothesized for the sand. Both laws are variations of the type 
           $D_{\rm max}/D_0 \propto We^{1/4}$. The pinkish shade emphasizes the area where the raindrop deforms itself but not frankly ``splashes'' since
           $D_{\rm max}/D_0 \lesssim 1$.}
\end{center}
\end{figure}

%
  A disdrometer is an instrument able to determine raindrop sizes and velocity distributions. In the absence of such a device 
on a planetary probe, one has to search for rainfall artifacts.
On the Earth, fresh raindrop imprints in dusty or
sandy terrains may be routinely observed. Fossil records have also been reported and used to constrain the air density existing billion years ago
\cite{som_etal_2012}, or to provide a way of retrodeforming plant or animal fossils \cite{fichman_2013}. An example of terrestrial fossil 
raindrop impressions is displayed in Fig.~\ref{RainDropImpre}. Extraterrestrial surfaces may also preserve more 
or less ancient traces of rainfalls.\\
    As a first approach, we restrict our discussion to an idealized scenario where a droplet hits a smooth and flat horizontal surface. Immediately after 
the impact, under the effect of inertia the liquid spreads circularly, this spreading is countered by the capillary and viscous forces
\cite{roisman_etal_2002}. We let $D_{\rm max}$ be the maximum diameter of the spreading, and $D_{0}$ the diameter of the droplet in flight. 
A universal scaling law, {\it i.e.} applicable to all regimes (capillary and viscous), may be found by introducing the impact 
parameter $P= We Re^{-2/5}$, where $We= \rho_{\rm liq} D_0 U^2 /\sigma$ is the Weber number which compares inertial and capillary forces. The quantity
$Re= \rho_{\rm liq} D_0 U /\eta$ represents the well known Reynolds number, accounting for the ratio between inertial and viscous 
forces. This law, allowing an excellent fit to empirical data, may be written as \cite{laan_etal_2014}
\begin{equation}\label{EqDmaxD0}
\frac{D_{\rm max}}{D_0} = Re^{1/5} \frac{P^{1/2}}{A + P^{1/2}}
\end{equation}
with $A$ a ``universal'' dimensionless constant, empirically determined: $A= 1.240 \pm 0.001$. Eq.~\ref{EqDmaxD0} 
connects the size distributions of raindrops and impacts
on a solid surface. In Fig.~\ref{FigDmaxD0} we have gathered results relevant for Titan,
the Earth and Mars. The raindrop minimum diameter $D_{\rm 0, min}$, considered in this figure, is determined by the condition 
$We \gtrsim 1$. Indeed, for $We < 1$ the raindrops are not really ``splashing'', but are rather gently deposited since the capillary forces dominate 
inertial ones, the deformation remains mild. The maximum diameters of in-flight raindrops $D_{\rm 0, max}$, have been fixed in accordance with Eq.~\ref{dmax}. 
All raindrop terminal velocities, required in $Re$ and $We$ evaluations, have been evaluated following the procedure described 
in Sec.~\ref{rainInducEros}.
The $D_{\rm max}/D_0$ appears to be an increasing function of $D_0$ (see Fig.~\ref{FigDmaxD0}), reflecting the growing dominant effect of 
inertia during impact, compared to the effect of surface tension. In the vicinity of $D_{\rm max}/D_0 = 1$ the fate of a landing 
raindrop is probably governed by capillary interaction with the soil and/or by evaporation (see Sec.~\ref{raindrop_abs}). In this region
the droplets deviate from the strict application of Eq.~\ref{EqDmaxD0}.
Moreover, for a fixed raindrop diameter the deformation during the impact is smaller on Mars (at least in the low pressure atmosphere) 
and Titan compared to what happens on the Earth. This is mainly caused by lower terminal velocities in the marsian and titanian cases.\\
    Up to this point, the rain-vector has been assumed to be perpendicular to the impacted surface, hypothesized  horizontal. 
The literature provides procedures to take into account inclined targeted surfaces \cite{laan_etal_2014}.\\
%
%
  More importantly, planetary surfaces are often covered by granular media: sand, regolith, sedimented aerosols, etc. These materials are known to 
exhibit behavior that cannot be approximated by a continuous solid. Then, the crater resulting from a raindrop impact on such granular 
medium should be different 
from what we get on a solid surface, at least in terms of spatial extension. The physics of the impact of a liquid droplet on a granular structure, 
is still the subject of active researches. Such a system is rather complex since it intricates the properties of the 
liquid and those of its targeted material; nonetheless a general behavior can be drawn. 
Assuming a full transfer of the drop-granular impact energy to the drop surface, leads to $D_{\rm max}/D_0 \propto We^{1/2}$
\cite{zhao_etal_2015}. However, recent works agree with a law of the type $D_{\rm max}/D_0 \propto We^{1/4}$ \cite{katsuragi_2010,delon_etal_2011,zhao_etal_2015}.
This power law is explained qualitatively by energy dissipation in internal degrees of freedom of the drop, and by the energy required for sand
excavation.
The effect of granular medium properties, such as the ratio of granular medium/water densities \cite{katsuragi_2010} or 
the packing fraction of the powder \cite{zhao_etal_2015} might be discussed but these details would not
change drastically the overall conclusion. In Fig.~\ref{FigDmaxD0}, we have 
plotted examples of $D_{\rm max}/D_0 \propto We^{1/4}$ laws (see the two lower curves) only in the case of Titan 
for clarity. As expected, raindrop spreading during impact on a 
sandy terrain is significantly smaller than in the case of a solid surface. Interestingly, the plotted laws converge to relatively mild 
deformation with a maximum around $D_{\rm max}/D_0 \simeq 3$. The average 
slopes of these curves are also smaller than those obtained for solid surface impacts, 
letting the raindrop size distribution less altered. This aspect is rather favorable to the use of sediment raindrop imprints, as a constraint on 
atmosphere microphysics.

\subsubsection{Application to Titan}
\label{TitanRainImpress}

   The {\it Cassini} spacecraft has revealed many evidences for the existence of a hydrological cycle
based on liquid methane on Titan. For instance, a lot of images of fluvial patterns have been acquired \cite{legall_etal_2010,langhans_etal_2012}, 
suggesting the occurrence of rainfall events in the past. Cloud activity has been clearly 
brought to light \cite{turtle_etal_2018}, but the observations of rainfalls have been scarce. Although such observations are more likely 
to be done in polar regions which appear to be the wettest, their in situ exploration belongs to the far future even if several mission
concepts have been already proposed 
(\textit{e.g.} \citeA{rodriguez_etal_2022}). 
Such observations are not excluded during the timespan of the
{\it Dragonfly} mission, which is dedicated to equatorial regions\rev{, namely relatively dry regions according to GCM models \cite{lora_etal_2019}}. 
Two episodes of rain showers have been
identified during the {\it Cassini} era: one in South polar regions \cite{turtle_etal_2009} and another in the equatorial zone \cite{turtle_eal_2011a}. 
In any case, the search for fossil raindrop imprints is perfectly relevant for {\it Dragonfly} 
which plans to investigate soils in environments rich in sediments \cite{lorenz_etal_2021}.
The impact of raindrops could happen on a flat solid surface in the natural surroundings of the probe, or even on its own 
hull; alternatively droplets may fall, or may have fallen, on a granular soil leaving more or less persistent traces. 
At the time of writing these words, we do not know if a flat and horizontal part of the {\it Dragonfly} fuselage will be accessible to camera fields of view. 
Beside clues concerning the raindrop size distribution based on $D_{\rm max}/D_0$ measurements, 
the observation of liquid drops deposited on a human made surface could bring information about the properties of the liquid, particularly via 
the contact angle determination. We emphasize that pre-flight empirical studies of probe surface wettability, by liquid methane (and other relevant
mixtures), could be very useful to in situ observations since low velocity droplet deposition and final state on the surface, depend on the surface wettability.
As shown in Fig.~\ref{FigDmaxD0}, liquid methane raindrop imprints in Titan sediments, recently formed or fossilized, may constitute interesting 
pieces of information on Titan processes which produce liquidometeors. The situation is all the more favorable as the drop sizes should be relatively 
little altered according to estimations reported in Fig.~\ref{FigDmaxD0}.
Flat geological surfaces, possibly covered by a thin layer of organic dust, could be the perfect target for the mentioned investigations. 
The existence of such targets is plausible in 
interdune regions which are planned to be explored by {\it Dragonfly} \cite{lorenz_etal_2021}. Such structures could be imaged by instruments 
belonging to the {\it Dragonfly Camera Suite} ({\it DragonCam} for short, \citeA{lorenz_etal_2018b})
that includes a microscopic imager able to characterize surface objects down to sand-grain scale.
\rev{Finally, we underline that hail may also reach the surface of Titan \cite{graves_etal_2008}, but observational evidences of such event 
are still not available. More generally,
other erosion processes are at work on Titan's surface, like eolian and chemical processes. Discussions may be found in the 
literature \cite{lorenz_lunine_1996}.}

\subsubsection{Application to Mars}

   In contrast to Titan, today Mars climate is particularly dry, leaving no hope for rainfall observations.
However, the Marsian weather may have undergone rainy episodes $\sim 3.6$ to $4$~billion years ago, and sedimentary 
deposits may have recorded artifacts of these events.\\
Orbital detection of clay stratigraphies on early Martian terrains is best explained if Mars experienced a period between the 
mid-Noachian ($> 3.85$~Gy) and the end of the Noachian ($\sim 3.7$~Gy) during which climatic conditions allowed persistent liquid 
water on its surface \cite{carter_etal_2015}. Rainfall may have occurred during this time period due to the resulting hydrological 
cycle of evaporation-precipitation, which is correctly modeled by the most recent 3D global climate simulations
\cite{turbet_forget_2021}. Therefore, the mudstone sediments formed on early Mars could have recorded 
raindrop imprints.\\
The Mars Science Laboratory (MSL) Curiosity rover landed in Gale crater on 6 August 2012. Gale crater formed 
near the time of the Noachian to Hesperian transition, $\sim 3.61$~Gy ago \cite{ledeit_etal_2013}. 
In situ observations indicate deposits attributed to a stream and a delta, marking the boundary of an ancient lake \cite{grotzinger_etal_2014}. 
Infilling began shortly after 
the crater formed and ended in the early Hesperian, likely $\sim 3.3$ to $3.2$~Gy \cite{grant_etal_2014}. The older outcrop explored 
by Curiosity rover, namely Sheepbed mustones, is interpreted as formed by settling of primary grains in a lake \cite{grotzinger_etal_2014}. 
It could have recorded raindrop imprints at a time when it was exposed to the atmosphere, as a coastal environment for instance, 
and to rainfall. However, the climatic conditions during the lake formation are important, as a too cold climate would result 
in snowfall instead of rainfall. Although morphological features indicate possible periglacial processes with a permafrost environment 
after the first hundred thousand years following the crater formation \cite{ledeit_etal_2013}, there is no evidence for a cold climate 
environment when the lake was present \cite{grotzinger_etal_2014}.
The Curiosity rover is equipped with a set of $17$ scientific and engineering cameras which
can provide complementary information on photographed raindrop imprints. 
The two Navigation Cameras (Navcam) acquire panoramas of the surroundings of the rover after each drive. They can resolve 
features of about $9$~mm in size at $2$~m. The front and rear Hazard Avoidance Cameras (Hazcam) take pictures on a daily basis and can 
resolve features down to $\sim 7$~mm in size. The Mast Camera (Mastcam) instrument is capable of resolving features of 
$\sim 2$~mm at a distance of $2$~m with its $34$~mm focal length lens, and $\sim 0.8$~mm at $2$~m with a lens of $100$~mm focal length
\cite{bell_etal_2017,malin_etal_2017}.
However, the field of view covered by the mosaics taken by these cameras is rarely as large as for the Navcam mosaics, except for the 
$360^{\rm o}$ Mastcam mosaics. The Remote Micro-Imager (RMI) is part of the ChemCam instrument aboard Curiosity rover. It can 
resolve features down to $\sim 200$~$\mu$m in size \cite{lemouelic_etal_2015} and is mainly used to document the geological context of the 
Laser-Induced Breakdown Spectrometer (LIBS) measurements. Finally, the Mars Hand Lens Imager (MAHLI) is a contact instrument
mounted on the turret at the end of Curiosity's 2 meter-long robotic arm 
\cite{edgett_etal_2012,edgett_etal_2015}.
MAHLI can resolve objects $45-60$~$\mu$m in size. The Curiosity rover has imaged hollow nodules on Sheepbed mudstone 
\cite{grotzinger_etal_2014}.
These are mm-scale circular rims with hollow centers. They have mean diameters of $1.2$~mm, and minimum/maximum diameters of $0.6/5.6$~mm. 
So far, the interpretation of these hollow nodules involves either the dissolution of the inner material, creating the void space, or gas 
bubbles formed during or soon after deposition of mudstone \cite{grotzinger_etal_2014}.\\
The Mars 2020 Perseverance rover landed in Jezero crater on 18 February 2021. Jezero crater formed during mid-Noachian, with an age 
greater than $3.82$~Gy \cite{sun_stack_2020}. Jezero crater has been interpreted as a paleolake 
\cite{fassett_head_2005}
containing a prominent fan-shaped body of sedimentary rocks deposited on its western margin \cite{schon_etal_2012}. 
On the basis of images taken by the Perseverance rover, the sedimentary deposits of this Gilbert-type delta indicate a transition from sustained 
hydrological activity in a persistent lake environment to highly energetic short-duration fluvial flows \cite{mangold_etal_2021}. 
The fluvio-lacustrine deposits, containing clay minerals \cite{horgan_etal_2020}, could have recorded raindrop imprints in the past.
Among the instruments available on Perseverance rover, the two Navcam cameras are well suited for raindrop imprints photography since they 
acquire panoramas of the surroundings of the rover after each drive and can resolve features $\sim 3.3$~mm at a distance of $2$~m. The Mastcam-Z, consisting in 
two identical cameras mounted on the Perseverance remote sensing mast, is capable of resolving features down to $\sim 0.7$~mm in size in the near 
field at the $110$~mm focal length setting.\\
As part of China's first Mars exploration mission ``Tianwen-1'', the Zhurong rover touched down on the surface of southern Utopia Planitia on 15 May 2021.
The Zhurong rover is equipped with a multispectral camera (MSCam) which is probably the most capable instrument for our purpose since it can 
resolve features $\sim 1.5$~mm in size. However, the region explored by the Zhurong rover has an age of $\sim 3.45$~Gy (Late Hesperian) 
\cite{wu_etal_2021}, 
when the Martian climate was very likely too cold for rainfalls. Furthermore, the landing area was resurfaced around the Middle Amazonian and the 
materials on the current terrain could be as young as $\sim 700$~My \cite{liu_etal_2022}
with no possible observations of raindrop impressions.

\section{Raindrop Absorption by a Porous Soil}
\label{raindrop_abs}

 On the Earth, the soil physics is an important topics \cite{lal_shukla_2004} since properties of soils have significant 
consequences on the water cycle. In extraterrestrial contexts, Titan is the only body where one could observe rainfall events in the 
next decades; thus we will restrict our discussion to this moon. The {\it Dragonfly} quadcopter is planned to explore a region over which 
large quantities of organic materials have sedimented.
The presence of porous terrains is therefore likely and precipitations of liquid methane (see Sec.~\ref{TitanRainImpress}) may be absorbed by the soil.  
%
\begin{figure}[!t]
\begin{center}
\includegraphics[angle=0, width=14 cm]{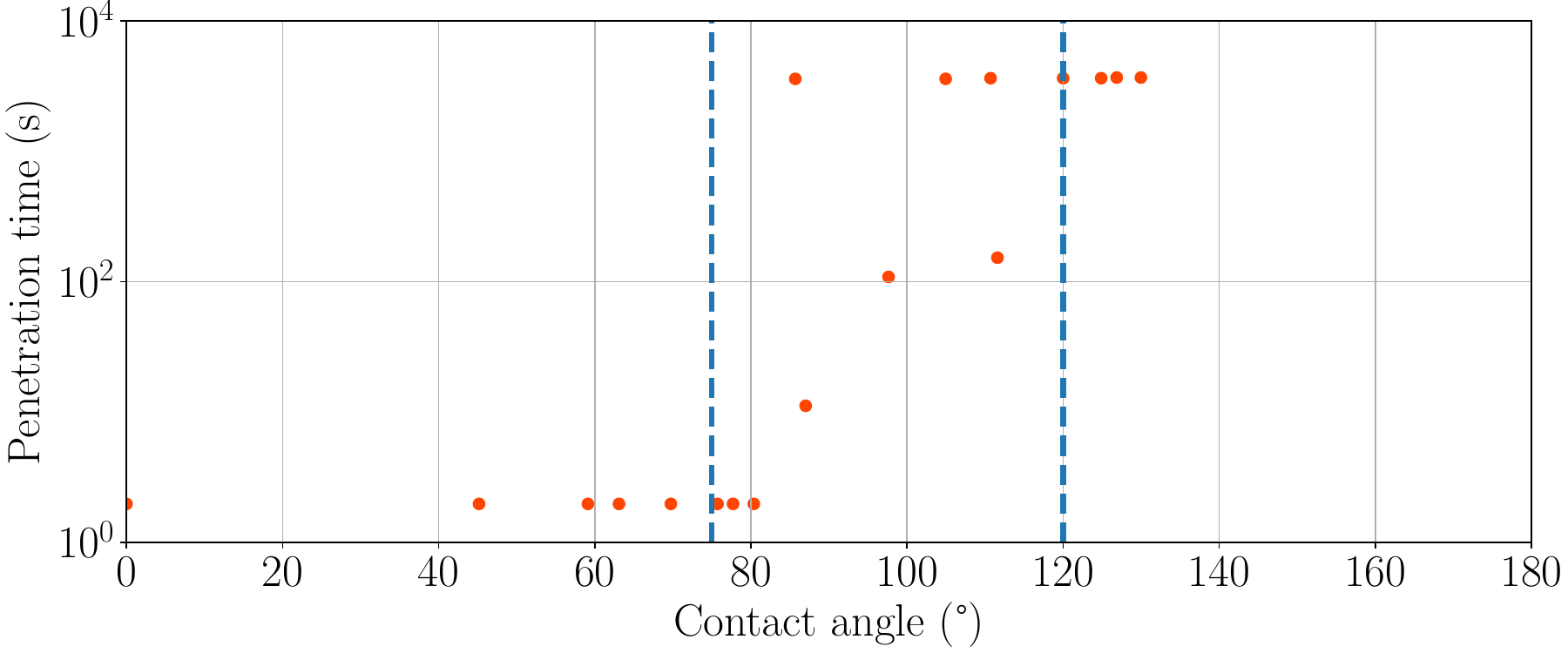}
\caption[]{\label{Fig6Bac03}Penetration time of water in a soil as a function of the contact angle, data read in Fig.~6 of 
           \citeA{bachmann_etal_2003}.}
\end{center}
\end{figure}
%
   As a first approach, a porous medium may be idealized as a large number of vertical parallel cylinders randomly distributed
\cite{hilpert_Ben-David_2009}. 
Right after its deposition on such a medium, a droplet of liquid starts to flow downward, 
filling up the ``tubes''. The temporal evolution of the length $h(t)$ of the liquid slug in ``tubes'' may be approached by the implicit 
non-linear equation \cite{washburn_1921,hilpert_Ben-David_2009}
\begin{equation}\label{hh0}
\frac{h}{h_0} - \frac{t}{t_0} = \mathrm{ln} \, \left(1 + \frac{h}{h_0}\right)
\end{equation}
with the length scale $h_0 = 2 \sigma \cos \theta_c / \rho g R_p$ and the time scale $t_0 = 16 \sigma \eta \cos \theta_c/\rho^2 g^2 R_p^3$
where $\eta$, $\sigma$, $\rho$, $g$ and $R_p$ are respectively the dynamic viscosity (Pa~s), the surface tension 
(N~m$^{-1}$), the density (kg~m$^{-3}$), the gravity (m~s$^{-2}$) and the radius (m) of the ``pores'' of the substrate. 
The numerical resolution of Eq.~\ref{hh0} gives results qualitatively comparable with $t_{\rm abs}$ \cite{hapgood_etal_2002}
\begin{equation}\label{tabs}
    t_{\rm abs} = 1.35 \frac{\eta V_d^{2/3}}{\sigma \epsilon^2 R_p \cos(\theta_c)}
\end{equation}
where $\epsilon$ is the porosity of the granular medium and $V_d$ the initial volume of a drop deposited on the considered medium.
In Eq.~\ref{tabs}, large porosity and pore radius favor short absorption times, while we have the opposite behavior by increasing the liquid
viscosity and the drop volume. Obviously, an equation like (\ref{tabs}) is valid only for liquidophilic substrates, and this absorption 
time gets longer for the most liquiphobic material since $t_{\rm abs} \propto 1/\cos(\theta_c)$. This tendency is confirmed by measurements
acquired on the Earth soils (see Fig.~\ref{Fig6Bac03}) where the absorption time is very long for liquidophobic materials.\\
  On Titan, if the deposition of raindrops on a porous soil occurs in the zone explored by {\it Dragonfly}, the observations could enable 
us to distinguish between liquidophilic and liquidophobic materials. 
The disappearance of raindrops should be ruled by droplet penetration in liquidophilic soils and droplet evaporation in liquidophobic soils.
The time scale of methane absorption due to the soil liquidophily, as derived from Eq.~(\ref{hh0}), is 
a fraction of second for a $2$~cm methane raindrop, whereas the 
evaporation should take $\sim 10$~hours according to empirical evaporation rates \cite{luspay_kuti_etal_2015}.
The type of observations
suggested here should be feasible with {\it Dragonfly} which is equipped with cameras and sensors measuring the air pressure, temperature,
wind speed and ``humidity'' (methane abundances) available in the {\it Dragonfly Geophysics and Meteorology Package} (DraGMet)
\cite{lorenz_etal_2018b}. The detection of ``methanophobic'' material sedimented at the surface of Titan is important due to its 
potential role in marine floating formation over polar seas \cite{cordier_carrasco_2019}.\\
  The investigations could be even pushed forwards if the soil porosity or the pore dimension are assessed by 
{\it Dragonfly} electromagnetic or optical measurements, since the numerical resolution of Eq.~\ref{hh0}
could lead to an estimation of the contact angle $\theta_c$. 

\begin{figure*}[!t]
\begin{center}
\includegraphics[angle=0, width=15 cm]{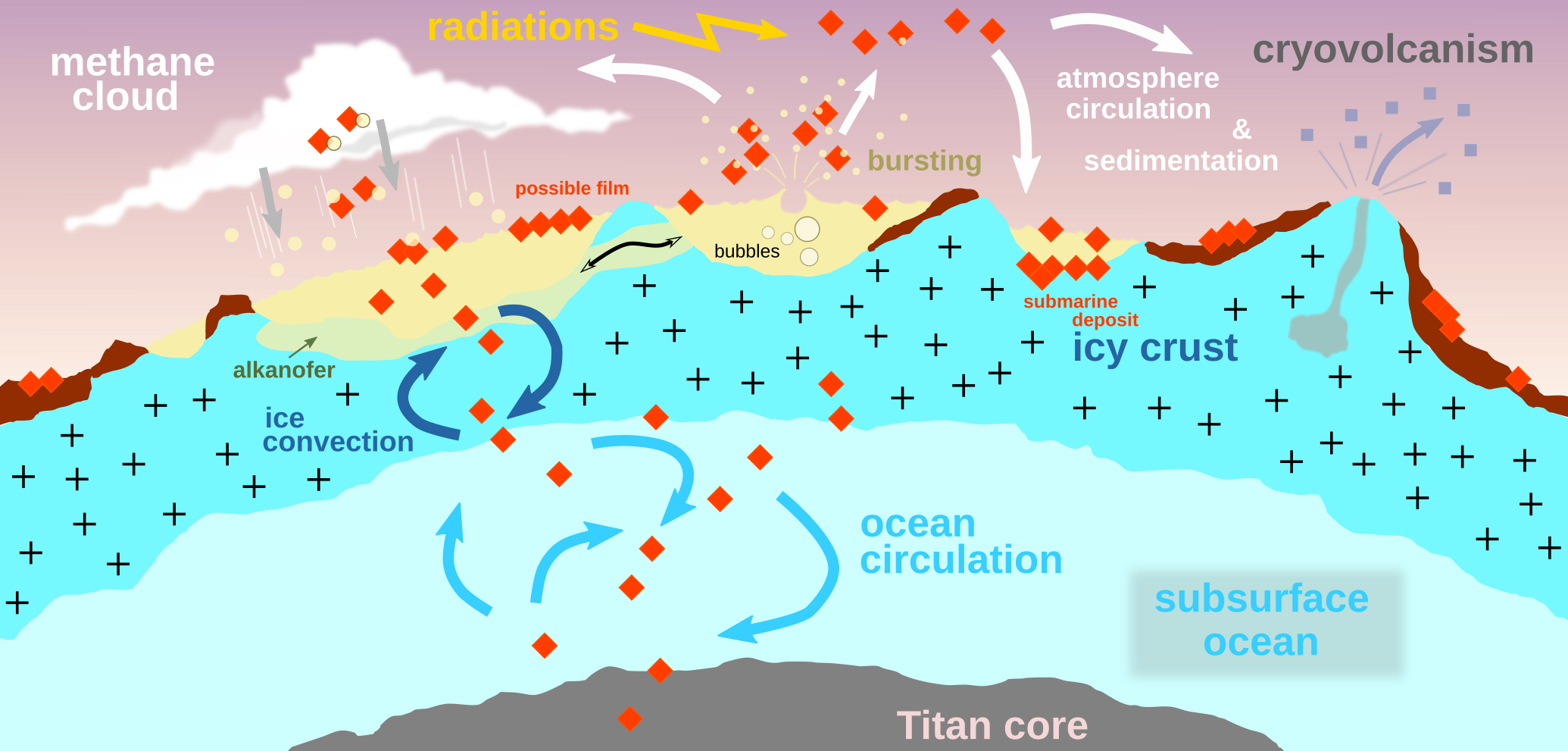}
\caption[]{\label{metacycle}A schematic view of a ``meta-cycle'' showing organic material travelling between the deepest regions 
           of Titan's interior, to the high altitude atmosphere, before going back to the moon core. The external part of nowadays Titan 
           core is probably made of high pressure water ice \cite{sotin_etal_2021}.
           For the sake of simplicity, the particles of organic material are represented by a unique type of orange diamond, but
           their nature can evolve during the course of their ``meta-cycle''.
           In each compartment organic matter is transported by the circulation of fluids, or semi-fluids in the case of icy crust. In a few
           environments, the material can be transformed, for instance by radiations in the atmosphere or by hydrolysis in the sub-surface 
           water ocean. In the atmosphere, organic particles can play the role of nucleation seeds for the formation of crystals or liquid droplets.
           The organic compounds can form floating films at the surface of lakes/seas, or sink to the bed, from which they could  be
           lifted off by marine circulation or bubbles streams. The bubble bursting may inject tiny droplets and organics in
           the air. The presence of alkanofers could ensure a form of communication between lakes.
           Finally, we have also represented massive organic layers at the surface (in brown), and the possible cryovolcanism that could 
           also provide the atmosphere with subsurface materials. The bubbling activity processes may have played a role in the emergence of molecular
           superstructures like azotosomes, one first step towards exobiological activity.}
\end{center}
\end{figure*}
%
\section{Conclusion}

   Given the incredible ubiquity of capillary processes at play in terrestrial environments, we have reviewed and discussed
some of them with possible importance in planetary contexts of our solar system. Having this common theme as a guideline, we have 
paid special attention to effects which could be directly or indirectly observed during near future space missions, 
or are of particular conceptual relevance. 
\rev{Of course we made choices and this work is not exhaustive, some other aspects could have been discussed, for instance the 
acoustic expression of bubble population in extraterrestrial waterfalls \cite{leighton_white_2004}, bubble formation due to
heat leak or cavitation around a boat or a submarine \cite{hartwig_etal_2016,cordier_2016}, or soil adhesion due to organic mud, in which 
capillarity induced the formation of liquid bridges between solid particles \cite{lorenz_2022}.}\\

  We have confirmed that the surface of Enceladus ocean is likely to be the most favorable 
place of our planetary system for organic matter flotation. As it is the case in numberless terrestrial situations, the ``bubble bursting''
is probably at the origin of the extraction, and concentration, of this material in tiny water droplets embedded in the flux of vapors 
ascending into geysers conduits. Concerning this ``bursting'', two distinct populations of bubbles can emerge at the interface 
between the liquid ocean and the vapor flux: 
(1) bubbles containing minors species (CO$_2$, CH$_4$, ...) coming from the deep interior, 
(2) bubbles of water vapor, formed at shallow depth with a size almost arbitrarily small when released from their nucleation site.
Since the very low gravity of Enceladus promotes large bubbles, and subsequently centimetric jet drops, the observed icy grains
are likely to have been produced either by particular phenomena as cascade bursting of large bubbles (of CH$_4$, CO$_2$, ...), 
or by bursting of tiny water vapor bubbles. The 
involved chemo-physical mechanisms have to be clarified; this is the reason why laboratory experiments of liquid water, close to $273$~K,
boiling under low pressure are highly desirable. Such investigations would be of particular interest in the perspective of 
{\it Europa Clipper} mission that could confirm the existence of geysers on Europa. This relevance is even reinforced in the case of the
future {\it in situ} exploration of Enceladus.\\
  
  The Titan's low gravity, combined with the nature of the working liquid, {\it i.e.} mainly liquid methane, favor bubbles and 
drops of sizes much larger than in the terrestrial context. The determination 
of size distributions would provide valuable information about the microphysics of the titanian low altitude atmospheric layers. Even if the areas which will be visited by {\it Dragonfly} are recognized to be rather dry, the observation of 
``fossil raindrop imprints'' is technically possible with {\it Dragonfly} instruments. 
Similar observations would be feasible on Mars, nonetheless they remain unlikely in view of the configuration of current Martian missions. The 
detection of raindrops, or dewdrops, on titanian ground would offer the opportunity 
to characterize soils ({\it e.g.}, the liquidophily of organic material sedimenting from the atmosphere).\\
As hypothesized in previous works, the hydrocarbon seas of Titan are prone to be covered, at least partially, by layers of organic materials 
sedimenting from the atmosphere. The nature of these layers may range from monomolecular to much thicker deposits. 
About organic grains, their size is a key parameter to assess their propensity to float.
Liquidophilic material could lead to floating matter provided that the particle size is small enough.
Although preliminary laboratory works have already been conducted, more in depth investigations are required since this
scientific question may have potential important consequences for Titan at global scale. Indeed, we have demonstrated (see Sec.~\ref{flapping_Titan}) 
that even in the absence of ``breaking waves'' at the surface of Titan's maria, the production of micronic aerosols may be driven by an aspect of bubble bursting called ``flapping shear instability'' which is likely to occur in the Titan's context.\\
   This bubble bursting specific mechanism is conceptually important since it makes possible the injection into the atmosphere of material 
previously processed in the interior of Titan. While cryovolcanism activity \cite{lopes_etal_2013} and cratering \cite{neish_lorenz_2012} could represent rare 
or very rare events, bubble bursting offers an 
alternate route to convey a broad variety of chemicals from the interior of Titan to its atmosphere.
The well identified maria and lacus are not the only sites where bursting could occur. The
transient presence of liquid hydrocarbons flushing the ground \cite{macKenzie_etal_2019,malaska_etal_2022} may be imagined in other regions, due to 
precipitations or dynamics of the local ``alkanofer'' \cite{cordier_etal_2021}. Fluxes of gases may, in such a situation, 
induce bubble-bursting, at the origin of the production of aerosols.\\
   Below the surface, a succession of transport phenomena could take turns with each other: the subsurface aqueous ocean ensures
the interaction with Titan core and the external icy crust \cite{sotin_etal_2021}, this ocean being surely well mixed by the convection. 
The deep core is probably a region rich in organic compounds, surrounded by a convecting high pressure ice shell. 
The convection in the icy crust could transport material from the ocean free surface up to a few kilometers above it \cite{kalousova_sotin_2020}. 
Finally, the liquid circulation in an ``alkanofer''
\cite{horvath_etal_2016,faulk_etal_2020,cordier_etal_2021} could release compounds directly to the surface or in a sea  
acting as an intermediate between the ``alkanofer'' and the atmosphere. Of course, matter transportation in the opposite direction, {\it i.e.}
from the atmosphere to the inner regions of Titan's interior, is easily conceivable. 
  Once put together, all these elements draw a global 
picture of a large ``meta-cycle'' made of a few ``environmental niches'' in interaction. Such a ``meta-cycle'' has already been advocated 
by Lerman with his theory of the ``Bubble-Aerosol-Droplet Cycle'', nicknamed ``bubblesol cycle'' in the context of the terrestrial emergence 
of life \cite{lerman_1986,lerman_teng_2005}. We propose an adaptation of this ``meta-cycle'' to Titan as sketched in Fig.~\ref{metacycle}.
This meta-cycle is an entanglement of several sub-cycles harbored in the atmosphere, the seas, the alkanofers, the crust and the water global 
ocean. In all these compartments, the transported material can be subject to chemo-physical changes. For instance, 
the organic substances, originally produced in the atmosphere, or coming from the deep core of Titan, could be altered during their
journey in the interior, undergoing hydrolysis in the aqueous ocean \cite{neish_etal_2009,neish_etal_2010}. Re-injected in the atmosphere by 
bubble-bursting and atomized in micronic particles, they can be transported downwind or higher in the atmosphere, and deposited in 
areas like those visited by {\it Dragonfly}.\\

   Up to this point, we have emphasized the role of bubble-bursting as a mechanism of aerosols production, seeding the whole
surface of Titan with material coming from the interior. However, bubble bursting has also two other effects that could be of paramount importance: (1) 
it can selectively desolvate organic compounds with mole fraction as low as 10$^{-9}$; for instance, the concentration of materials dissolved in water 
was found to be increased by a factor of $10^4$ after desolvatation \cite{lerman_teng_2005};
(2) in presence of surface active molecules, 
the production of tiny droplets promotes the formation of bilayer vesicles which are prerequisites for the emergence of life similar to the 
terrestrial ones \cite{chang_1993,schulze-makuch_irwin_2018}.\\
 The effect of selectivity and concentration of compounds in droplets produced by bursting is well known when the solvent is water, and
similar effect can be expected with a cryogenic solvent mainly based on methane. Once in the air, each micronic droplet can behave like 
a micro-reactor in which chemical reactions are favored. Energy deposition on the droplet may even accelerate processes.\\
 All forms of life we know on the Earth are compartmentalized: multicellular organisms are, by definition, composed 
 of several entities bounded by a membrane, bacteria are structured by a molecular wall (the cytoplasmic membrane) while equivalently viruses have their 
capsids. The self-emergence of these ``molecular walls'' may have been favored, or catalyzed, by the presence of bubbles in the 
primordial seas. These ideas have already been discussed decades ago \cite{chang_1993} in works according to which bubble formation, 
and bursting at the surface, may be seen as a plausible mechanism for closing vesicles. The production of synthetic liposomes, based on 
bubbling, has also been proposed by the pharmaceutical industry \cite{talsma_etal_1994}. The appearance of membrane promoted by bubbles may be 
advocated in both Titan- and Enceladus-like environments.

     The ``meta-cycle'', proposed for Titan, can be also envisaged for Enceladus where a part of the material blowed off by geysers 
is deposited over enceladean surfaces \cite{robidel_etal_2020}. The considered layer may be subsequently transported back to the ocean with 
the help of the icy crust convection. More localized sub-cycles may be devised. Inside cracks, right above the water surface, bursting could 
favor the formation of vesicles and higher concentrations, which will finally go back to the ocean surface.\\

     In this work, we have focused our discussion on solar system objects whose liquid phases will be available
to observation and exploration in upcoming decades. During the beginning of its operations, the JWST has already 
shown the possible existence of water on the exoplanet WASP-96~b, while the hunting
for exomoons is at its premises \cite{kipping_etal_2022}. The detection of stable liquid phases on these classes of objects is
a promising field \cite{maltagliati_2019,trees_stam_2019}. However, detailed studies of capillarity effects in such objects 
will remain a distant dream for a long time yet.

\section*{Data Availability Statement}

The computer software developed for this work is publicly available in the Zenodo archive
\cite{cordier_2023}.

\acknowledgments
We gratefully acknowledge the Reviewers Dr.~Ralph Lorenz and Pr.~Jason Barnes for their very constructive suggestions that improved significantly
our manuscript.
We thank Gabriel Tobie and Benjamin Charnay for scientific discussion, and  Michael Rygel for sharing his fossil raindrop imprint picture.
The present research was supported by the Programme National de Plan\'{e}tologie (PNP) of CNRS-INSU co-funded by CNES, and also
partially supported by the HPC center of Champagne-Ardenne (ROMEO).
This work has been done mainly using open-source softwares like \texttt{Python}, \texttt{gfortran}, \texttt{kate}, \texttt{Jupyter Notebook},
\LaTeX{ }under \texttt{GNU/Debian Linux} operating system. The authors warmly acknowledge 
the whole Free Software community.

%
%
%
\bibliographystyle{agujournal2019}


\def\sciam{Sci.
  Am.}\def\nature{Nature}\def\nat{Nature}\def\science{Science}\def\natastro{Nat.
  Astron.}\def\natgeo{Nat. Geosci.}\def\natcom{Nat.
  Commun.}\def\pnas{PNAS}\def\AnnderPhys{‎Ann. Phys.
  (Berl.)}\def\icarus{Icarus}\def\pss{Planet. Space
  Sci.}\def\psj{PSJ}\def\planss{Planet. Space Sci.}\def\ssr{Space Sci.
  Rev.}\def\solsr{Sol. Syst. Res.}\def\ApSSP{ApSSP}\def\psj{Planet. Sci.
  J.}\def\jqsrt{J. Quant. Spectrosc. Radiat. Transfer}\def\expastro{Exp.
  Astron.}\def\jcis{‎J. Colloid Interface
  Sci.}\def\aap{A\&A}\def\apj{ApJ}\def\apjl{ApJL}\def\apjs{ApJS}\def\aj{AJ}\def\mnras{MNRAS}\def\araa{Annu.
  Rev. Astron. Astrophys.}\def\areps{Annu. Rev. Earth Planet.
  Sci.}\def\pasj{Publ. Astron. Soc. Jpn.}\def\apss{Astrophys. Space
  Sci.}\def\expastro{Exp. Astron.}\def\pasp{Publ. Astron. Soc.
  Pac.}\def\expastron{Exp. Astron.}\def\asr{Adv. Space
  Res.}\def\galaxies{Galaxies}\def\astrobiol{Astrobiology}\def\areps{Annu. Rev.
  Earth Planet. Sci.}\def\georl{Geophys. Res. Lett.}\def\grl{Geophys. Res.
  Lett.}\def\jgr{J. Geophys. Res.}\def\gca{Geochim. Cosmochim.
  Ac.}\def\epsl{Earth Planet. Sci. Lett.}\def\ess{Earth Space
  Sci.}\def\plasci{Planet. Sci.}\def\ggg{Geochem. Geophys.
  Geosyst.}\def\rmg{Rev. Mineral. Geochem.}\def\gji{Geophys. J.
  Int.}\def\tpm{Transport Porous Med.}\def\philtrans{Phil.
  Trans.}\def\faradis{Farad. Discuss.}\def\jcis{‎J. Colloid Interface
  Sci.}\def\jfm{J. Fluid Mech.}\def\physflu{Phys. Fluids}\def\pachem{Pure Appl.
  Chem.}\def\jpcA{J. Phys. Chem. A}\def\chemrev{Chem. Rev.}\def\AppOpt{Appl.
  Opt.}\def\nature{Nature}\def\nat{Nature}\def\science{Sci}\def\pnas{Proc.
  Natl. Acad. Sci. U.S.A.}\def\jced{J. Chem. Eng. Data}\def\fpe{Fluid Phase
  Equilibria}\def\iecr{Ind. Eng. Chem. Res.}\def\aichej{AIChE J.}\def\pt{Powder
  Technol.}\def\etfs{Exp. Therm. Fluid Sci.}\def\mineng{Miner.
  Eng.}\def\intjminprocess{Int. J. Miner. Process}\def\jcolintersci{J. Colloid
  Interface
  Sci.}\def\nature{Nature}\def\nat{Nature}\def\science{Sci}\def\jced{J. Chem.
  Eng. Data}\def\fpe{Fluid Phase Equilibria}\def\iecr{Ind. Eng. Chem.
  Res.}\def\aichej{AIChE J.}\def\pt{Powder Technol.}\def\etfs{Exp. Therm. Fluid
  Sci.}\def\jgr{J. Geophys. Res.}\def\jcp{J. Chem. Phys.}\def\jcis{‎J.
  Colloid Interface Sci.}\def\jcsft{J. Chem. Soc. Faraday
  Trans.}\def\nature{Nature}\def\nat{Nature}\def\science{Science}\def\natastro{Nat.
  Astron.}\def\natgeo{Nat. Geosci.}\def\natcom{Nat. Commun.}\def\scirep{Sci.
  Rep.}\def\sciad{Sci.
  Adv.}\def\aap{A\&A}\def\apj{ApJ}\def\apjl{ApJL}\def\apjs{ApJS}\def\aj{AJ}\def\mnras{MNRAS}\def\araa{Annu.
  Rev. Astron. Astrophys.}\def\pasj{Publ. Astron. Soc.
  Jpn.}\def\apss{Astrophys. Space Sci.}\def\pasp{Publ. Astron. Soc.
  Pac.}\def\expastron{Exp. Astron.}\def\asr{Adv. Space
  Res.}\def\galaxies{Galaxies}\def\science{Sci}\def\jced{J. Chem. Eng.
  Data}\def\fpe{Fluid Phase Equilibria}\def\iecr{Ind. Eng. Chem.
  Res.}\def\aichej{AIChE J.}\def\pt{Powder Technol.}\def\etfs{Exp. Therm. Fluid
  Sci.}\def\jgr{J. Geophys. Res.}\def\gca{Geochim. Cosmochim.
  Acta}\def\chemgeol{Chem Geol.}\def\jcp{J. Chem. Phys.}\def\jcis{‎J. Colloid
  Interface Sci.}\def\jcsft{J. Chem. Soc. Faraday Trans.}\def\jpcB{J. Phys.
  Chem. B}\def\jsf{J. Supercrit. Fluids}\def\enerp{Energy
  Procedia}\def\aichej{AlChE J.}\def\IECPDD{Ind. Eng. Chem. Process Des.
  Dev.}\def\EF{Energy Fuels}\def\jacs{J. Am. Chem. Soc.}\def\sciam{Sci.
  Am.}\def\nature{Nature}\def\nat{Nature}\def\science{Science}\def\natastro{Nat.
  Astron.}\def\natgeo{Nat. Geosci.}\def\natcom{Nat. Commun.}\def\sciad{Sci.
  Adv.}\def\scirep{Sci. Rep.}\def\AnnderPhys{‎Ann. Phys.
  (Berl.)}\def\icarus{Icarus}\def\pss{Planet. Space Sci.}\def\planss{Planet.
  Space Sci.}\def\ssr{Space Sci. Rev.}\def\solsr{Sol. Syst.
  Res.}\def\expastro{Exp. Astron.}\def\jcis{‎J. Colloid Interface
  Sci.}\def\aap{A\&A}\def\apj{ApJ}\def\apjl{ApJL}\def\apjs{ApJS}\def\aj{AJ}\def\mnras{MNRAS}\def\aapr{A\&ARv}\def\araa{Annu.
  Rev. Astron. Astrophys.}\def\pasj{Publ. Astron. Soc.
  Jpn.}\def\apss{Astrophys. Space Sci.}\def\pasp{Publ. Astron. Soc.
  Pac.}\def\expastron{Exp. Astron.}\def\astrobiol{Astrobiology}\def\areps{Annu.
  Rev. Earth Planet. Sci.}\def\georl{Geophys. Res. Lett.}\def\jgr{J. Geophys.
  Res.}\def\gca{Geochim. Cosmochim. Ac.}\def\epsl{Earth Planet. Sci.
  Lett.}\def\plasci{Planet. Sci.}\def\ggg{Geochem. Geophys.
  Geosyst.}\def\tpm{Transport Porous Med.}\def\philtrans{Phil.
  Trans.}\def\faradis{Farad. Discuss.}\def\jcis{‎J. Colloid Interface
  Sci.}\def\jfm{J. Fluid Mech.}\def\physflu{Phys.
  Fluids}\def\pnas{PNAS}\def\pachem{Pure Appl. Chem.}\def\jpcA{J. Phys. Chem.
  A}\def\chemrev{Chem.
  Rev.}\def\nature{Nature}\def\nat{Nature}\def\science{Science}\def\natastro{Nat.
  Astron.}\def\natgeo{Nat. Geosci.}\def\natcom{Nat. Commun.}\def\scirep{Sci.
  Rep.}\def\sciad{Sci. Adv.}\def\pnas{Proc. Natl. Acad. Sci.
  }\def\aap{A\&A}\def\apj{ApJ}\def\apjl{ApJL}\def\apjs{ApJS}\def\aj{AJ}\def\mnras{MNRAS}\def\araa{Annu.
  Rev. Astron. Astrophys.}\def\pasj{Publ. Astron. Soc.
  Jpn.}\def\apss{Astrophys. Space Sci.}\def\pasp{Publ. Astron. Soc.
  Pac.}\def\expastron{Exp. Astron.}\def\asr{Adv. Space
  Res.}\def\galaxies{Galaxies}\def\science{Sci}\def\jced{J. Chem. Eng.
  Data}\def\fpe{Fluid Phase Equilibria}\def\iecr{Ind. Eng. Chem.
  Res.}\def\aichej{AIChE J.}\def\pt{Powder Technol.}\def\etfs{Exp. Therm. Fluid
  Sci.}\def\jgr{J. Geophys. Res.}\def\gca{Geochim. Cosmochim.
  Acta}\def\chemgeol{Chem Geol.}\def\jcp{J. Chem. Phys.}\def\jcis{‎J. Colloid
  Interface Sci.}\def\jcsft{J. Chem. Soc. Faraday Trans.}\def\jpcB{J. Phys.
  Chem. B}\def\jsf{J. Supercrit. Fluids}\def\enerp{Energy
  Procedia}\def\aichej{AlChE J.}\def\IECPDD{Ind. Eng. Chem. Process Des.
  Dev.}\def\EF{Energy Fuels}\def\jacs{J. Am. Chem. Soc.}\def\sciam{Sci.
  Am.}\def\nature{Nature}\def\nat{Nature}\def\science{Science}\def\natastro{Nat.
  Astron.}\def\natgeo{Nat. Geosci.}\def\natcom{Nat.
  Commun.}\def\pnas{PNAS}\def\AnnderPhys{‎Ann. Phys.
  (Berl.)}\def\icarus{Icarus}\def\pss{Planet. Space
  Sci.}\def\psj{PSJ}\def\planss{Planet. Space Sci.}\def\ssr{Space Sci.
  Rev.}\def\solsr{Sol. Syst. Res.}\def\ApSSP{ApSSP}\def\psj{Planet. Sci.
  J.}\def\jqsrt{J. Quant. Spectrosc. Radiat. Transfer}\def\expastro{Exp.
  Astron.}\def\jcis{‎J. Colloid Interface
  Sci.}\def\aap{A\&A}\def\apj{ApJ}\def\apjl{ApJL}\def\apjs{ApJS}\def\aj{AJ}\def\mnras{MNRAS}\def\araa{Annu.
  Rev. Astron. Astrophys.}\def\pasj{Publ. Astron. Soc.
  Jpn.}\def\apss{Astrophys. Space Sci.}\def\pasp{Publ. Astron. Soc.
  Pac.}\def\expastron{Exp. Astron.}\def\asr{Adv. Space
  Res.}\def\galaxies{Galaxies}\def\astrobiol{Astrobiology}\def\areps{Annu. Rev.
  Earth Planet. Sci.}\def\georl{Geophys. Res. Lett.}\def\grl{Geophys. Res.
  Lett.}\def\jgr{J. Geophys. Res.}\def\jgrp{J. Geophys. Res.
  Planets}\def\gca{Geochim. Cosmochim. Ac.}\def\epsl{Earth Planet. Sci.
  Lett.}\def\ess{Earth Space Sci.}\def\plasci{Planet. Sci.}\def\ggg{Geochem.
  Geophys. Geosyst.}\def\rmg{Rev. Mineral. Geochem.}\def\gji{Geophys. J.
  Int.}\def\tpm{Transport Porous Med.}\def\philtrans{Phil.
  Trans.}\def\faradis{Farad. Discuss.}\def\jcis{‎J. Colloid Interface
  Sci.}\def\jfm{J. Fluid Mech.}\def\physflu{Phys. Fluids}\def\pachem{Pure Appl.
  Chem.}\def\jpcA{J. Phys. Chem. A}\def\chemrev{Chem. Rev.}\def\AppOpt{Appl.
  Opt.}\def\nature{Nature}\def\nat{Nature}\def\science{Science}\def\natastro{Nat.
  Astron.}\def\natphys{Nat. Phys.}\def\natgeo{Nat. Geosci.}\def\natcom{Nat.
  Commun.}\def\scirep{Sci. Rep.}\def\science{Sci}\def\jced{J. Chem. Eng.
  Data}\def\fpe{Fluid Phase Equilibria}\def\iecr{Ind. Eng. Chem.
  Res.}\def\aichej{AIChE J.}\def\pt{Powder Technol.}\def\etfs{Exp. Therm. Fluid
  Sci.}\def\jgr{J. Geophys. Res.}\def\grl{Geophys. Res. Lett.}\def\gca{Geochim.
  Cosmochim. Acta}\def\jcp{J. Chem. Phys.}\def\jpcl{J. Phys. Chem.
  Lett.}\def\jcis{‎J. Colloid Interface Sci.}\def\jcsft{J. Chem. Soc. Faraday
  Trans.}\def\jpcB{J. Phys. Chem. B}\def\jsf{J. Supercrit.
  Fluids}\def\enerp{Energy Procedia}\def\aichej{AlChE J.}\def\IECPDD{Ind. Eng.
  Chem. Process Des. Dev.}\def\pre{Phys. Rev. E.}\def\orggeoch{Org. Geochem.}



\appendix
\section{Conditions for floatability by capillarity}
\label{AppCondi}
%
\begin{figure}[]
\begin{center}
\includegraphics[angle=0, width=8 cm]{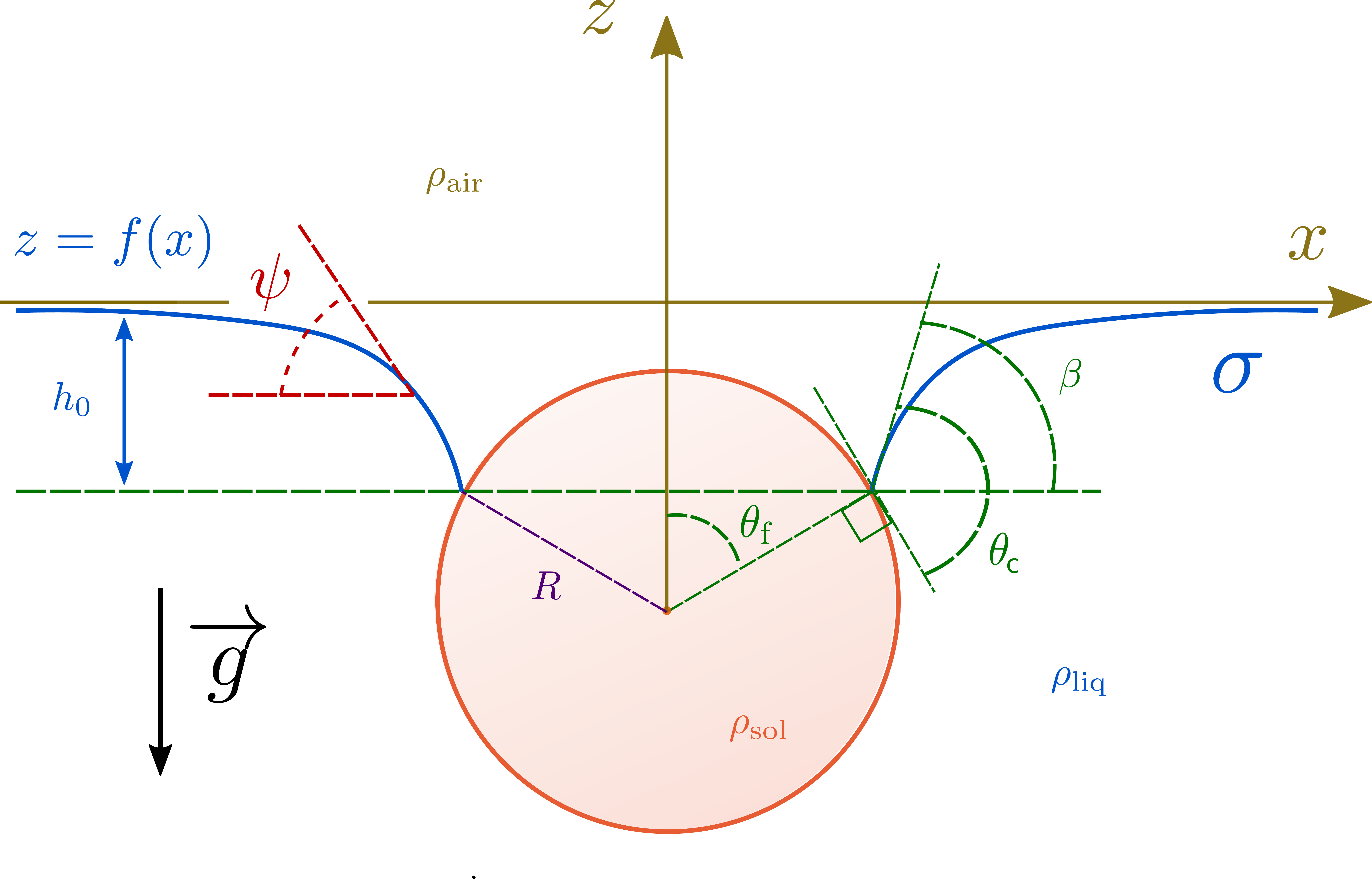}
\caption[]{\label{Fig1Lee18}Sketch of a spherical solid particle, of radius $R$ and density $\rho_{\rm sol}$, floating at the surface
           of a liquid with density $\rho_{\rm liq}$. The meniscus (blue line) is represented by the equation $z=f(x)$, the surface
           tension of the interface is $\sigma$, while the contact angle between the 
           solid and the liquid is denoted $\theta_c$; $\theta_{\rm f}$ is the filling angle. The difference of height between
           the plane of flotation and the surface of the unperturbed liquid is $h_0$. The local tangent to the meniscus has the angle
           $\psi$ with the horizontal direction. The gravity is represented by $\overrightarrow{g}$.
           This figure is our reconstruction of Fig.~1 published by \citeA{lee_2018}, it is also very similar to Fig.~2 in
           \citeA{scheludko_etal_1976}. }
\end{center}
\end{figure}
%
   In this frame, let us define a spherical solid particles; of radius $R$ 
and density $\rho_{\rm sol}$. This object is assumed to be deposited at the surface of a liquid characterized by a density $\rho_{\rm liq}$ 
and a surface tension $\sigma$ (see Fig.~\ref{Fig1Lee18}). According to the literature \cite{lee_2018}, this particle undergoes three forces. 
The most obvious of them, is the weight
\begin{equation}
   F_M = \frac{4\pi}{3} \rho_{\rm sol} R^3 g
\end{equation}
The second one is the capillary force, whose vertical component $F_C$ has the expression
\begin{equation}
   F_C = 2\pi \, R \, \sin \theta_f \, (\sigma \sin \beta)
\end{equation}
where $\theta_f$ is the filling angle and $\beta$ includes both the contact angle $\theta_c$
and $\theta_f$ (see Fig.~\ref{Fig1Lee18}: $\beta=\theta_c-\theta_f$). Taking into account forces caused by air and liquid, coming 
from the classical Archimede's principle, and also generated by liquid above the line of flotation ({\it i.e.} corresponding to the 
``layer'' of thickness $h_0$ in Fig.~\ref{Fig1Lee18}), the resulting buoyancy force is 
\begin{equation}
\begin{split}
  F_B = &    \frac{4\pi}{3} R^3 \rho_{\rm air} g 
           + \frac{\pi}{3} R^3 (2 + 3 \cos \theta_f - \cos^3 \theta_f) \times \\
        & (\rho_{\rm liq} - \rho_{\rm air}) g
          - \pi R^2 \sin^2\theta_f \, h_0 \, (\rho_{\rm liq} - \rho_{\rm air}) \, g
\end{split}
\end{equation}
The flotation of the particle is reached at equilibrium when we have the force balance 
\begin{equation}\label{balance}
        F_M= F_C + F_B
\end{equation}
Unfortunately, this equation cannot be solved alone since it contains two unknowns: the filling angle $\theta_f$ and $h_0$; other parameters, 
particularly the contact angle $\theta_c$, specify the nature of the system, and are assumed to be given. As a requirement, we have to include
the Young-Laplace equation
\begin{equation}
    \kappa_{m} = \frac{\rho_{\rm liq} - \rho_{\rm air}}{2 \sigma} g z
\end{equation}
which governs the profile of the meniscus (blue line in Fig.~\ref{Fig1Lee18}).  In this equation, $\kappa_{m}$ is the local mean curvature given
by the non-linear differential equation
\begin{equation}\label{mean_curvature}
  \kappa_{m} = -\frac{1}{2} \, \left( \frac{f''}{(1+f'^2)^{3/2}} + \frac{f'}{x (1 + f'^2)^{1/2}} \right)
\end{equation}
where $z= f(x)$ is the equation of the meniscus profile. According to the dedicated literature \cite{lee_2018}, the system of equations 
(\ref{balance})-(\ref{mean_curvature}) can be reformulated under the form of a pair of differential equations, solved numerically using 
a shooting method \cite{lee_2018}. In Fig.~\ref{influParam}~(a) we show an example of computed menisci profiles, illustrating the influence
of the density ratio $D = (\rho_{\rm sol} - \rho_{\rm air}) / (\rho_{\rm liq} - \rho_{\rm air})$.
 By expanding and reformulating Eq.~\ref{balance}, we can write
\begin{equation}\label{reformulated}
\frac{g}{6\sigma}( \rho_{\rm liq} -\rho_{\rm air}) R^2 
    = \frac{\sin \theta_f \, \sin (\theta_c - \theta_f) }{4D - (2 + 3 \cos\theta_f - \cos^3 \theta_f)
             + 3\frac{h_0}{R} \sin^2\theta_f}
\end{equation}
Since, in the present context, the density of the solid is larger than the liquid one, the ratio $D$ remains larger than 1 
(a typical value is around $\sim 2$); in addition, we have $ -2 \le 3\cos\theta_f - \cos^3 \theta_f \le +2$ for all the permitted 
$\theta_f$ values, which are in the range $0^{\rm o}-180^{\rm o}$. As a consequence, the denominator of the right-hand side of 
Eq.~\ref{reformulated} is positive. Then, the liquid being denser than the air, $\sin(\theta_c - \theta_f)$ has to be positive and, 
importantly, the filling angle $\theta_f$ has to remain smaller than the contact angle $\theta_c$.\\
   For $\theta_f \lesssim 45^{\rm o}$, the term $3\cos\theta_f - \cos^3 \theta_f$ remains reasonably close to $2$, and 
$3 h_0/R \sin^2 \theta_f$ is negligible, in such a case
\begin{equation}
\frac{g}{6\sigma}(\rho_{\rm liq} -\rho_{\rm air}) R^2 \simeq
     \frac{\sin \theta_f \, \sin (\theta_c - \theta_f) }{4D - 4}
\end{equation}
It can be easily shown \cite{scheludko_etal_1976,crawford_ralston_1988} that the radius has a maximum value 
$R_{\rm max}$ for $\theta_f = \theta_c/2$ and
\begin{equation}\label{Rmax}
  R_{\rm max} \simeq \sqrt{\frac{3\sigma}{2 (\rho_{\rm sol} - \rho_{\rm liq}) g}} \, \sin \frac{\theta_c}{2}
\end{equation}


%
%
%
%
%

\end{document}